\documentclass[12pt,onecolumn]{IEEEtran}
\usepackage{amsmath}
\usepackage{amssymb}
\usepackage{cite}
\usepackage{color}
\usepackage{epsfig}
\usepackage{epsf}
\usepackage{rotating}

\renewcommand{\baselinestretch}{1.2}
\newtheorem{thm}{Theorem}
\newtheorem{lemma}{Lemma}
\newtheorem{Corollary}{Corollary}

\begin{document}

\thispagestyle{empty} \vskip 1cm

\begin{center}
\vskip 0.6cm
{\large \bf Communication Over MIMO Broadcast Channels Using Lattice-Basis Reduction}\footnote{This work was supported in part by funding from Communications and Information Technology Ontario (CITO), Nortel Networks, and Natural Sciences
and Engineering Research Council of Canada (NSERC). The material of this paper was presented at the IEEE International
Symposium on Information Theory, Adelaide, Australia, September 2005.}  \\

\vskip 0.8cm Mahmoud Taherzadeh, Amin Mobasher, and Amir K. Khandani

\vskip 0.5cm

{\small Coding \& Signal Transmission Laboratory\\
Department of Electrical \& Computer Engineering\\
University of Waterloo\\
Waterloo, Ontario, Canada, N2L 3G1\\
}

\end{center}

%_________________________________________________________
\begin{abstract}
A simple scheme for communication over MIMO broadcast channels is
introduced which adopts the lattice reduction technique to improve
the naive channel inversion method. Lattice basis reduction helps us
to reduce the average transmitted energy by modifying the region
which includes the constellation points. Simulation results show
that the proposed scheme performs well, and as compared to the more
complex methods (such as the perturbation method \cite{PeelHoch})
has a negligible loss. Moreover, the proposed method is extended to
the case of different rates for different users. The asymptotic behavior (SNR$\longrightarrow
\infty $) of the symbol error rate of the proposed method and the perturbation technique, and also the outage probability for the
case of fixed-rate users is analyzed. It is shown that the proposed
method, based on LLL lattice reduction, achieves the optimum asymptotic slope of symbol-error-rate
(called the precoding diversity). Also, the outage probability for the
case of fixed sum-rate is analyzed.

\end{abstract}

\section{Introduction}

In the recent years, communications over multiple-antenna fading
channels has attracted the attention of many researchers. Initially,
the main interest has been on the point-to-point Multiple-Input
Multiple-Output (MIMO) communications \cite{Tel99, Fos96, Tarokh98, TJ99, Alamouti}. In \cite{Tel99} and
\cite{Fos96}, the authors have shown that the capacity of a MIMO
point-to-point channel increases  linearly with the minimum number
of the transmit and the receive antennas.

More recently, new information theoretic results \cite{caire-shamai},
\cite{YuCioffi}, \cite{ViswanathTse}, \cite{VJG} have shown that in
multiuser MIMO systems, one can exploit most of the advantages of
multiple-antenna systems. It has been shown that in a MIMO broadcast
system, the sum-capacity grows linearly with the minimum number of
the transmit and receive antennas \cite{YuCioffi},
\cite{ViswanathTse}, \cite{VJG}. To achieve the sum capacity, some information theoretic schemes,
based on dirty-paper coding, are introduced. Dirty-paper coding was
originally proposed for the Gaussian interference channel when the
interfering signal is known at the transmitter \cite{Costa}. Some
methods, such as using nested lattices, are introduced as
practical techniques to achieve the sum-capacity promised by the
dirty-paper coding \cite{Erez}. However, these methods are not easy to
implement.

As a simple precoding scheme for MIMO broadcast systems, the channel
inversion technique (or zero-forcing beamforming
\cite{caire-shamai}) can be used at the transmitter to separate the
data for different users. To improve the performance of the channel
inversion technique, a zero-forcing approximation of the dirty paper
coding (based on QR decomposition) is introduced in
\cite{caire-shamai} (which can be seen as a scalar approximation of \cite{Erez}). However, both of these methods are vulnerable
to the poor channel conditions, due to the occasional near-singularity
of the channel matrix (when the channel matrix has at least one small eigenvalue). This
drawback results in a poor performance in terms of the
symbol-error-rate for the mentioned methods \cite{PeelHoch}.

In \cite{PeelHoch}, the authors have introduced a \textit{vector
perturbation technique} which has a good performance in terms of
symbol error rate. Nonetheless, this technique requires a lattice
decoder which is an NP-hard problem. To reduce the complexity of the
lattice decoder, in \cite{fischer2003, fischer, fischer2004, fischer2006}, the authors have used
lattice-basis reduction to approximate the closest lattice point
(using Babai approximation).

In this paper, we present a transmission technique for the MIMO
broadcast channel based on the lattice-basis reduction. Instead of
approximating the closest lattice point in the perturbation
problem, we use the lattice-basis reduction to reduce the average
transmitted energy by reducing the second moment of the
fundamental region generated by the lattice basis. This viewpoint
helps us to: (i) achieve a better performance as compared to
\cite{fischer}, (ii) expand the idea for the case of unequal-rate
transmission, and (iii) obtain some analytic results for the
asymptotic behavior (SNR$\longrightarrow \infty $) of the
symbol-error-rate for both the proposed technique and the
perturbation technique of \cite{PeelHoch}.

The rest of the paper is organized as the following: Sections II and III briefly describe the system model and introduce
the concept of lattice basis reduction. In section IV, the proposed
method is described and in section V, the proposed approach is
extended for the case of unequal-rate transmission. In section VI, we consider the asymptotic performance of the
proposed method for high SNR values, in terms of the probability of error.
We define the precoding diversity and the outage probability for the
case of fixed-rate users. It is shown that by using lattice
basis reduction, we can achieve the maximum precoding diversity. For
the proof, we use a bound on the orthogonal deficiency of an
LLL-reduced basis. Also, an upper bound is given for the probability
that the length of the shortest vector of a lattice (generated by
complex Gaussian vectors) is smaller than a given value. Using
this result, we also show that the perturbation technique achieves
the maximum precoding diversity. In section VII, some simulation results are presented. These
results show that the proposed method offers almost the same
performance as \cite{PeelHoch} with a much smaller complexity. As
compared to \cite{fischer}, the proposed method offers almost the same performance. However, by sending a very small amount of side information (a few bits for one fading block), the modified proposed method offers a better
performance with a similar complexity. Finally, in section VIII,
some concluding remarks are presented.

\section{System Model and Problem Formulation}

We consider a multiple-antenna broadcast system with $N_{t} $ transmit antennas and $N_{r} $ single-antenna users ($N_{t}\geq N_{r}$).
Consider $ \mathbf{y}=[y_{1},...,y_{N_{r}}]^{T} $, $
\mathbf{x}=[x_{1},...,x_{N_{t}}]^{T} $, $
\mathbf{w}=[w_{1},...,w_{N_{r}}]^{T} $, and the $ N_{r}\times N_{t} $ matrix
$\mathbf{H}$, respectively, as the received signal, the transmitted
signal, the noise vector, and the channel matrix. The transmission
over the channel can be formulated as,
\begin{equation}
\mathbf{y}=\mathbf{H}\mathbf{x}+\mathbf{w} .
\end{equation}

The channel is assumed to be Raleigh, i.e. the elements of
$\mathbf{H}$ are i.i.d. with the zero-mean unit-variance complex
Gaussian distribution and the noise is i.i.d. additive Gaussian. Moreover, we
have the energy constraint on the transmitted signal, $ \textmd{E} (
\Vert \mathbf{x} \Vert^{2} ) =1$. The energy of the additive noise is
$ \sigma^{2} $ per antenna, i.e. $ \textmd{E}  ( \Vert \mathbf{w}
\Vert^{2} ) =N_{r}\sigma^{2} $. The Signal-to-Noise Ratio
(SNR) is defined as $ \rho=\frac{1}{\sigma^{2}} $.

In a broadcast system, the receivers do not cooperate with each
other (they should decode their respective data, independently). The
main strategy in dealing with this restriction is to apply an
appropriate precoding scheme at the transmitter. The simplest method
in this category is using the channel inversion technique at the
transmitter to separate the data for different users:
\begin{equation}
\mathbf{s}=\mathbf{H}^{+}\mathbf{u},
\end{equation}
where $
\mathbf{H}^{+}=\mathbf{H}^{\mathtt{H}}(\mathbf{H}\mathbf{H}^{\mathtt{H}})^{-1}
$, and $ \mathbf{H}^{\mathtt{H}}$ is the Hermitian of $\mathbf{H} $.
Moreover, $ \mathbf{s} $ is the transmitted signal before the
normalization ($ \mathbf{x}=\dfrac{\mathbf{s}}{ \sqrt{\textmd{E}  (
\Vert \mathbf{s} \Vert^{2} )}}$ is the normalized transmitted
signal), and $ \mathbf{u} $ is the data vector, i.e. $ u_{i} $ is
the data for the \textit{i}'th user. For $N_{t}= N_{r}$ (the number of transmit
antennas and the number of users are equal), the transmitted signal is
\begin{equation}
\mathbf{s}=\mathbf{H}^{-1}\mathbf{u}.
\end{equation}

The problem arises when $\mathbf{H}$ is poorly conditioned and $ \Vert
\mathbf{s} \Vert $ becomes very large, resulting in a high power
consumption. This situation occurs when at least one of the
singular values of $\mathbf{H}$ is very small which results in vectors
with large norms as the columns of $\mathbf{H}^{+}$. Fortunately, most of the time
(especially for high SNRs), we can combat the effect of a small
singular value by changing the supporting region of the constellation which is the main motivation behind the current article.

When the data of different users are selected from $ \mathbb{Z}[i]$, the overall constellation can be seen as a set of lattice
points. In this case, lattice algorithms can be used to modify the constellation. Especially, lattice-basis reduction is a
natural solution for modifying the supporting region of the
constellation.

\section{Lattice-Basis Reduction}

Lattice structures have been frequently used in different
communication applications such as quantization or decoding of MIMO
systems. A real (or complex) lattice $ \Lambda $ is a discrete set
of $ N $-D vectors in the real Euclidean space $ \mathbb{R}^{N} $
(or the complex Euclidean space $ \mathbb{C}^{N} $) that forms a
group under ordinary vector addition. Every lattice $ \Lambda $ is
generated by the integer linear combinations of some set of linearly
independent vectors $ \mathbf{b}_{1},\cdots ,\mathbf{b}_{M} \in
\Lambda $, where the integer $ M$, $M\leq N $, is called the
dimension of the lattice $ \Lambda $.
The set of vectors $ \left\lbrace \mathbf{b}_{1},\cdots ,\mathbf{b}_{M} \right\rbrace $ is
called a basis of $ \Lambda $, and the matrix $ \mathbf{B} = \left[ \mathbf{b}_{1},
\cdots , \mathbf{b}_{M}\right]  $, which has the basis vectors as its
columns, is called the basis matrix (or generator matrix) of $
\Lambda $.

The basis for representing a lattice is not unique. Usually a basis consisting of relatively short and nearly orthogonal
vectors is desirable. The procedure of finding such a basis for a
lattice is called \textit{Lattice Basis Reduction}. A popular
criterion for lattice-basis reduction is to find a basis such that $ \Vert \mathbf{b}_{1} \Vert \cdot...\cdot \Vert \mathbf{b}_{M}
\Vert $ is minimized. Because the volume of the lattice\footnote{Volume of the lattice generated by $ \mathbf{B}$ is $\left( \det \mathbf{B}^{\mathtt{H}}\mathbf{B}\right) ^{\frac{1}{2}} $.} does not change with the change of basis, this problem is equivalent to minimizing the orthogonality defect which is defined as
\begin{equation} \label{eq:OrthD}
\delta \triangleq \frac{(\Vert \mathbf{b}_{1}\Vert^{2} \Vert \mathbf{b}_{2}\Vert^{2}...\Vert
\mathbf{b}_{M}\Vert^{2})}{\det \mathbf{B}^{\mathtt{H}}\mathbf{B}
}.
\end{equation}

The problem of finding such a basis is NP-hard \cite{Lbook}.
Several distinct sub-optimal reductions have been studied in the
literature, including those associated to the names Minkowski,
Korkin-Zolotarev, and more recently Lenstra-Lenstra and Lovasz (LLL)
\cite{lekkerkerker}.

An ordered basis $ \left( \mathbf{b}_{1}, \cdots , \mathbf{b}_{M}
\right) $ is a \emph{Minkowski-Reduced Basis} \cite{Hel85} if
\begin{itemize}
\item $ \mathbf{b}_{1} $ is the shortest nonzero vector in the
lattice $ \Lambda $, and \item For each $k=2,...,M$, $ \mathbf{b}_{k}
$ is the shortest nonzero vector in $ \Lambda $ such that $
\left( \mathbf{b}_{1}, \cdots , \mathbf{b}_{k} \right)$
may be extended to a basis of $ \Lambda $.
\end{itemize}
Minkowski reduction can be seen as a greedy solution for the lattice-basis reduction problem. However, finding Minkowski reduced basis is equivalent to finding the
shortest vector in the lattice and this problem by itself is
NP-hard. Thus, there is no polynomial time algorithm for this
reduction method.

In \cite{LLL82}, a reduction algorithm (called \textit{LLL algorithm}) is introduced which uses the Gram-Schmidt orthogonalization and has a polynomial complexity and guarantees a bounded orthogonality defect. For any ordered
basis of $ \Lambda $, say $ \left( \mathbf{b}_{1},\cdots ,\mathbf{b}_{M} \right) $, one
can compute an ordered set of Gram-Schmidt vectors, $
\left( \hat{\mathbf{b}}_{1},\cdots ,\hat{\mathbf{b}}_{M} \right) $, which are
mutually orthogonal, using the following recursion:

\begin{equation}
\begin{tabular}{c}
$ \hat{\mathbf{b}}_{i}=\mathbf{b}_{i} - \sum_{j=1}^{i-1}\mu
_{ij}\hat{\mathbf{b}}_{j}, \textmd{ with} $\\
$\mu_{ij}= \dfrac{< \mathbf{b}_{i}, \hat{\mathbf{b}}_{j}>}
{<\mathbf{b}_{j}, \hat{\mathbf{b}}_{j}>}. $
\end{tabular}
\end{equation}
where $<\cdot, \cdot> $ is the inner product.
An ordered basis $ \left( \mathbf{b}_{1}, \cdots ,
\mathbf{b}_{M} \right) $ is an \emph{LLL Reduced Basis}
\cite{LLL82} if,
\begin{itemize}
\item $ \Vert \mu_{ij} \Vert \leq \frac{1}{2} $ for $ 1 \leq i < j \leq M $, and
\item $ p \cdot \Vert \hat{\mathbf{b}}_i \Vert^2 \leq  \Vert \hat{\mathbf{b}}_{i+1} + \mu_{i+1,i}\hat{\mathbf{b}}_i \Vert^2 $
\end{itemize}
where $\frac{1}{4}< p <1$, and $ \left( \hat{\mathbf{b}}_{1}, \cdots ,
\hat{\mathbf{b}}_{M} \right) $ is the Gram-Schmidt
orthogonalization of the ordered basis and $
\mathbf{b}_{i}=\sum_{j=1}^{i}\mu_{ij}\hat{\mathbf{b}}_{j} $ for $
i=1,...,M$. 

It is shown that LLL basis-reduction algorithm produces relatively
short basis vectors with a polynomial-time computational complexity
\cite{LLL82}. The LLL basis reduction has found extended applications in
several contexts due to its polynomial-time complexity. In \cite{Napias}, LLL algorithm is generalized for Euclidean rings (including the ring of complex integers). In this
paper, we will use the following important property of the complex LLL
reduction (for $p=\frac{3}{4} $):

\begin{thm} [see \cite{Napias}]
Let $\Lambda$ be an $ M $-dimensional complex lattice and $\mathbf{ B} =[\mathbf{b}_{1}...\mathbf{b}_{M}]   $ be the LLL reduced basis of $
\Lambda $. If $ \delta$ is the orthogonality defect of $ \mathbf{B}$, then,
\begin{equation} \sqrt{\delta} \leq  2^{M(M-1)} .
\end{equation}
\end{thm}

\section{Proposed Approach}

Assume that the data for different users, $ u_i $, is selected
from the points of the integer lattice (or from the half-integer
grid \cite{Forney1}). The data vector $ \mathbf{u} $ is a point in
the Cartesian product of these sub-constellations. As a result,
the overall receive constellation consists of the points from
$\mathbb{Z}^{2N_{r}} $, bounded within a $2N_{r}$-dimensional hypercube.
At the transmitter side, when we use the channel inversion
technique, the transmitted signal is a point inside a
parallelotope whose edges are parallel to vectors, defined by the
columns of $ \mathbf{H}^{+} $. If the data is a point from the
integer lattice $ \mathbb{Z}^{2N_{r}} $, the transmitted signal is a
point in the lattice generated by $ \mathbf{H}^{+} $. When the
squared norm of at least one of the columns of $ \mathbf{H}^{+} $
is too large, some of the constellation points require high energy
for the transmission. We try to reduce the average transmitted
energy, by replacing these points with some other points with
smaller square norms. However, the lack of cooperation among the
users imposes the restriction that the received signals should
belong to the integer lattice $ \mathbb{Z}^{2N_{r}} $ (to avoid the
interference among the users). The core of the idea in this paper
is based on using an appropriate supporting region for the
transmitted signal set to minimize the average energy, without
changing the underlying lattice. This is achieved through the
lattice-basis reduction.

When we use the continuous approximation (which is appropriate for
large constellations), the average energy of the transmitted signal
is approximated by the second moment of the transmitted region \cite{Forney1}. When
we assume equal rates for the users, e.g. $ R $ bits per user
($\frac{R}{2}$ bits per dimension), the signal points (at the receiver) are inside a hypercube with an edge of length $a $ where
\begin{equation}\label{eq:aR}
a= 2^{R/2} .
\end{equation}
Therefore, the supporting region of the transmitted signal is the
scaled version of the fundamental region of the lattice generated
by $ \mathbf{H}^{+}$ (corresponding to its basis) with the
scaling factor $ a $. Note that by changing the basis for this
lattice, we can change the corresponding fundamental region (a
parallelotope generated by the basis of the lattice and centered at
the origin). The second moment of the resulting region is
proportional to the sum of the squared norms of the basis vectors
(see Appendix A). Therefore, we should try to find a basis
reduction method which minimizes the sum of the squared norms of
the basis vectors. Figure 1 shows the application of the lattice
basis reduction in reducing the average energy by replacing the
old basis with a new basis which has shorter vectors. In this
figure, by changing the basis $\mathbf{a}_{1},\mathbf{a}_{2}$
(columns of $ \mathbf{H}^{+}$) to $\mathbf{b}_{1},\mathbf{b}_{2}$
(the reduced basis), the fundamental region $\mathcal{F} $,
generated by the original basis, is replaced by
$\mathcal{F}^{\prime} $, generated by the reduced basis.

\begin{figure}
  \centering
  \includegraphics[scale=.6]{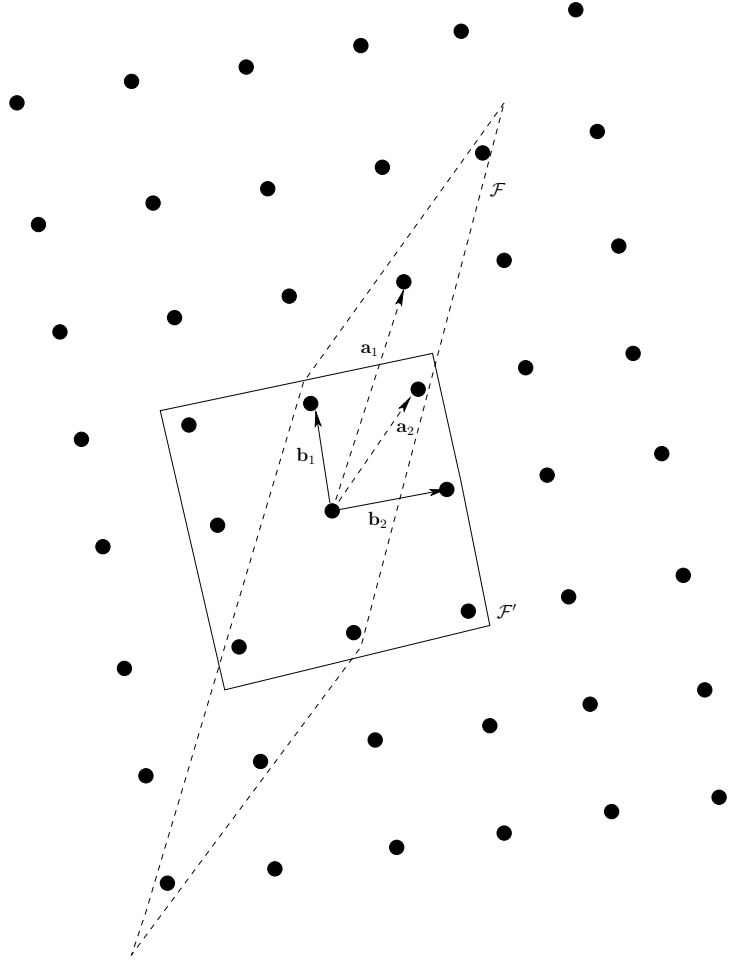}
\vspace{30pt}

  \caption{Using lattice-basis reduction for reducing the average energy}

\end{figure}

Among the known reduction algorithms, the Minkowski reduction can
be considered as an appropriate greedy algorithm for our problem.
Indeed, the Minkowski algorithm is the successively optimum solution
because in each step, it finds the shortest vector. However, the
complexity of the Minkowski reduction is equal to the complexity of
the shortest-lattice-vector problem which is known to be NP-hard \cite{Ajtai98}. Therefore, we use the LLL
reduction algorithm which is a suboptimum solution with a polynomial
complexity.

Assume that $ \mathbf{B}=\mathbf{H}^{+}\mathbf{U} $ is the
LLL-reduced basis for the lattice obtained by $ \mathbf{H}^{+} $,
where $ \mathbf{U}$ is an $N_{r} \times N_{r}$ unimodular matrix (both $\mathbf{U}$ and
$\mathbf{U}^{-1}$ have integer entries). We use $
\mathbf{x}=\mathbf{Bu}^{\prime}=\mathbf{H}^{+}\mathbf{U}
\mathbf{u}^{\prime} $ as the transmitted signal where
\begin{equation}\label{eq:transmod}
\mathbf{u}^{\prime}= \mathbf{U}^{-1} \mathbf{u} \mod a
\end{equation}
is the precoded data vector, $\mathbf{u} $ is the original data
vector, and $ a$ is the length of the edges of the hypercube, defined by (\ref{eq:aR}). At the receiver side, we use modulo operation to find the
original data:

\begin{equation}\label{eq:8}
\begin{tabular}{c}
$ \mathbf{y}=\mathbf{Hx}+\mathbf{n}=
\mathbf{HH}^{+} \mathbf{U}(\mathbf{U}^{-1}\mathbf{u} \mod a) +\mathbf{n} =\mathbf{U}(\mathbf{U}^{-1}\mathbf{u} \mod a)+\mathbf{n}$
\end{tabular}
\end{equation}
\begin{equation} \label{eq:9}
=\mathbf{UU}^{-1}\mathbf{u} \mod a+\mathbf{n}
=\mathbf{u} \mod
a+\mathbf{n}. 
\end{equation}
In obtaining (\ref{eq:9}) from (\ref{eq:8}), we use the fact that $\mathbf{U} $ and $ \mathbf{U}^{-1}$ have integer entries.

In this method, at the beginning of each fading block, we reduce
the lattice obtained by $ \mathbf{H}^{+} $ and during this block
the transmitted signal is computed using (\ref{eq:transmod}).
Neglecting the preprocessing at the beginning of the block (for
lattice reduction), the complexity of the precoding is in the
order of a matrix multiplication and a modulo operation.
Therefore, the complexity of the proposed precoding method is
comparable to the complexity of the channel inversion method.
However, as we will show by the simulation results, the
performance of this method is significantly better, and indeed, is
near the performance of the perturbation method, presented in
\cite{PeelHoch}.

In the perturbation technique \cite{PeelHoch}, the idea of
changing the support region of the constellation has been
implemented using a different approach. In \cite{PeelHoch}, $
\mathbf{u}^{\prime}=\mathbf{u}+a\mathbf{l}$ is used as the
precoded data, where the integer vector $\mathbf{l}$ is chosen to
minimize $\Vert \mathbf{H}^{+}(\mathbf{u}+a\mathbf{l}) \Vert$.
This problem is equivalent to the closest-lattice-point problem
for the lattice generated by $ a\mathbf{H}^{+}$ (i.e. finding the lattice point which is closer to $-
\mathbf{H}^{+}\mathbf{u}$). Therefore, in the perturbation
technique, the support region of the constellation is a scaled
version of the Voronoi region \cite{Sloan} of the lattice. In the
proposed method, we use a parallelotope (generated by the reduced
basis of the lattice), instead of the Voronoi region.  Although
this approximation results in a larger second moment (i.e. higher
energy consumption), it enables us to use a simple precoding
technique, instead of solving the closest-lattice-point problem.

For the lattice constellations, using a parallelotope instead of
the Voronoi region (presented in this paper) is equivalent with
using the Babai approximation instead of the exact lattice
decoding (previously presented in \cite{fischer}). The only practical difference between our lattice-reduction-aided scheme and the scheme presented in \cite{fischer} is that we reduce $\mathbf{H}^{+}$, while in \cite{fischer}, $\mathbf{H}^{\mathtt{H}}$ is reduced. As can be observed in the simulation results (figure 2), this difference has no significant effect on the performance. However, the new viewpoint
helps us in extending the proposed method for the case of
variable-rate transmission and obtaining some analytical results
for the asymptotic performance.

The performance of the proposed lattice-reduction aided scheme can be improved by combining it with other schemes, such as regularization \cite{PeelHoch1} or the V-BLAST precoding \cite{Joham}, or by sending a very small amount
of side information. In the rest of this section, we present two of these modifications. 

\subsection{Regularized lattice-reduction-aided precoding}

In \cite{PeelHoch1}, the authors have proposed a regularization scheme to reduce the transmitted power, by avoiding the near-singularity of $\mathbf{H} $. In this method, instead of using $\mathbf{H}^{+} =\mathbf{H}^{\mathtt{H}}\left( \mathbf{H}\mathbf{H}^{\mathtt{H}}\right)^{-1}$, the transmitted vector is constructed as 
\begin{equation}
\mathbf{x}=\mathbf{H}^{\mathtt{H}}\left( \mathbf{H}\mathbf{H}^{\mathtt{H}}+\alpha \mathbf{I}\right)^{-1} \mathbf{u}
\end{equation}
where $ \alpha$ is a positive number. To combine the regularized scheme with our lattice-reduction-aided scheme, we consider $ \mathbf{B}_{r}$ as the matrix corresponding to the reduced basis of the lattice generated by $\mathbf{H}^{\mathtt{H}}\left( \mathbf{H}\mathbf{H}^{\mathtt{H}}+\alpha \mathbf{I}\right)^{-1} $. When we use the regularization, the received signals of different users are not orthogonal anymore and the interference acts like extra noise. Parameter $\alpha$ should be optimized such that the ratio between the power of received signal and the power of the effective noise is minimized \cite{PeelHoch1}.

\subsection{Modified lattice-reduction-aided precoding with small side information}
In practical systems, we are interested in using a subset of
points with odd coordinates from the integer lattice. In these cases, we can improve the
performance of the proposed method by sending a very small amount
of side information. When the data vector $ \mathbf{u} $ consists
of odd integers, using the lattice-basis
reduction may result in points with some even coordinates (i.e.
$\mathbf{U}^{-1} \mathbf{u}$ has some even elements), instead of
points with all-odd coordinates in the new basis. For this case, in (\ref{eq:transmod}), the set of precoded data $ \mathbf{u}^{\prime}$ is not centered at the origin, hence
the transmitted constellation (which includes all the valid points $\mathbf{B} \mathbf{u}^{\prime}$) is not centered at the origin. Therefore, we can
reduce the transmitted energy and improve the performance by
shifting the center of the constellation to the origin. It can be
shown that the translation vector is equal to $ (\mathbf{U}^{-1}
[1+i, 1+i,\cdots ,1+i]^{\mathtt{T}} +[1+i, 1+i,\cdots ,1+i]^{\mathtt{T}}) \mod 2$ where $ i=\sqrt{-1}$.
When we use this shifted version of the constellation, we must
send the translation vector to the users (by sending 2 bits per
user) at the beginning of the block. However, compared to the
size of the block of data, the overhead of these two bits is
negligible.

The above idea of using a shift vector can be also used to improve
the perturbation technique (if we only use the odd points of the
lattice). After reducing the inverse of the channel matrix and
obtaining the bits (corresponding to the shift vector) at the
beginning of each fading block, the closest point to the signal
computed in equation (\ref{eq:transmod}) can be found by using the
sphere decoder. Then, the transmitted signal is obtained by
\begin{equation} \label{eq:modpert}
\mathbf{x}=\mathbf{B}\left( \mathbf{u}^{\prime} + a\mathbf{l} +
\mathbf{u}_{par} \right),
\end{equation}
where $ \mathbf{u}_{par} $ is the zero-one shift vector, which is computed
for users at the beginning of the fading block, and the perturbation
vector $ \mathbf{l} $ is an even integer vector such that the vector
$ \mathbf{x} $ has the minimum energy. This method which can be considered as \textit{modified perturbation
method} outperforms the perturbation method in \cite{PeelHoch}. When we are not restricted to the odd lattice points, using (\ref{eq:modpert}) instead of $ \mathbf{H}^{+}\left( \mathbf{u}^{\prime} + a\mathbf{l} \right)$ does not change the performance of the perturbation method. It only reduces the complexity of the lattice decoder \cite{agrell}.

\section{Unequal-Rate Transmission}

In the previous section, we had considered the case that the transmission
rates for different users are equal. In some applications, we are interested in assigning different rates to different users. 
Consider $ R_{1}, ..., R_{N_{r}}$ as the transmission rates for the users (we consider them as even integer numbers). Equation (\ref{eq:transmod}) should be modified as 
\begin{equation}\label{eq:transmodUn}
\mathbf{u}^{\prime}= \mathbf{U}^{-1} \mathbf{u} \mod \mathbf{a},
\end{equation}
where the entries of $\mathbf{a}=[a_{1},...,a_{N_{r}}]^{T} $ are equal to
\begin{equation}\label{eq:aUn}
a_{i}= 2^{R_{i}/2} .
\end{equation}
Also, at the receiver side, instead of (\ref{eq:8}) and (\ref{eq:9}), we have
\begin{equation}\label{eq:8Un}
\begin{tabular}{c}
$ \mathbf{y}=\mathbf{Hx}+\mathbf{n}=
\mathbf{HH}^{+} \mathbf{U}(\mathbf{U}^{-1}\mathbf{u} \mod \mathbf{a}) +\mathbf{n} =\mathbf{U}(\mathbf{U}^{-1}\mathbf{u} \mod \mathbf{a})+\mathbf{n}$
\end{tabular}
\end{equation}
\begin{equation} \label{eq:9Un}
=\mathbf{UU}^{-1}\mathbf{u} \mod \mathbf{a}+\mathbf{n}
=\mathbf{u} \mod
\mathbf{a}+\mathbf{n}. 
\end{equation}

If we are interested in
sum-rate, instead of individual rates, we can improve the
performance of the proposed method by assigning variable rates to different users. We assume that the sum-rate (rather than the
individual rates) is fixed and we want to reduce the average
transmitted energy. To simplify the analysis, we use the continuous
approximation which has a good accuracy for high rates.

Considering continuous approximation, the sum-rate is proportional
to the logarithm of the volume of the lattice with basis $
\mathbf{B} $ and the average energy is proportional to the second moment of the corresponding prallelotope, which is proportional to $
\sum_{i=1}^{N_{r}} \Vert \mathbf{b}_{i} \Vert^{2}= \textmd{tr}\mathbf{B}\mathbf{B}^{\mathtt{H}} $ (see Appendix A). The goal is to minimize the
average energy while the sum-rate is fixed. We can use another
lattice generated by $ \mathbf{B}^{\prime} $ with the same volume,
where its basis vectors are scaled versions of the vectors of the
basis $ \mathbf{B} $, according to different rates for different
users. Therefore, we can use $ \mathbf{B}^{\prime}=\mathbf{BD}$
instead of $\mathbf{B}$ (where $\mathbf{D} $ is a unit determinant $N_{r}\times N_{r} $
diagonal matrix which does not change the volume of the lattice).
For a given reduced basis $ \mathbf{B} $, the product of the
squared norms of the new basis vectors is constant:
\begin{equation}
\begin{tabular}{c}
$ \Vert \mathbf{b}_{1}^{\prime}\Vert^{2} \Vert
\mathbf{b}_{2}^{\prime}\Vert^{2}...\Vert
\mathbf{b}_{N_{r}}^{\prime}\Vert^{2}=(\Vert \mathbf{b}_{1}\Vert^{2}
\Vert \mathbf{b}_{2}\Vert^{2}...\Vert
\mathbf{b}_{N_{r}}\Vert^{2}) \det \mathbf{D} $\\
$ = \Vert \mathbf{b}_{1}\Vert^{2} \Vert
\mathbf{b}_{2}\Vert^{2}...\Vert \mathbf{b}_{N_{r}}\Vert^{2}={\rm
const}.$
\end{tabular}
\end{equation}

The average energy corresponding to the new lattice basis should be
minimized. When we use the modified basis $ \mathbf{B}^{\prime}$
instead of $\mathbf{B} $, the average
energy is proportional to $ \sum_{i=1}^{N_{r}} \Vert
\mathbf{b}_{i}^{\prime} \Vert^{2}= \textmd{tr}
\mathbf{B}^{\prime}\mathbf{B}^{\prime\mathtt{H}} $ (see Appendix A). According to the
arithmetic-geometric mean inequality, $ \sum_{i=1}^{N_{r}} \Vert
\mathbf{b}_{i}^{\prime} \Vert^{2}= \textmd{tr}
\mathbf{B}^{\prime}\mathbf{B}^{\prime\mathtt{H}} $ is minimized iff
\begin{equation}\label{eq:unequal}
\Vert \mathbf{b}_{1}^{\prime}\Vert= \Vert \mathbf{b}_{2}^{\prime}
\Vert=...=\Vert \mathbf{b}_{N_{r}}^{\prime}\Vert.
\end{equation}
Therefore,
\begin{equation}\label{eq:unequal2}
\min \textmd{tr}\mathbf{B}^{\prime}\mathbf{B}^{\prime\mathtt{H}}= N_{r} \left(  \Vert \mathbf{b}_{1}^{\prime}\Vert^{2} \Vert
\mathbf{b}_{2}^{\prime}\Vert^{2}...\Vert \mathbf{b}_{N_{r}}^{\prime}\Vert^{2} \right) ^{\frac{1}{N_{r}}} =  N_{r} \left(  \Vert \mathbf{b}_{1}\Vert^{2} \Vert
\mathbf{b}_{2}\Vert^{2}...\Vert \mathbf{b}_{N_{r}}\Vert^{2} \right) ^{\frac{1}{N_{r}}}
\end{equation}

Having the matrix $ \mathbf{B} $, the columns of matrix $
\mathbf{B}^{\prime} $ can be found using the equation
(\ref{eq:unequal}) and $ \textmd{tr}\mathbf{B}^{\prime}\mathbf{B}^{\prime\mathtt{H}}$ can be obtained by (\ref{eq:unequal2}). Now, for the selection of the reduced basis $ \mathbf{B}$, we
should find $ \mathbf{B} $ such that $\Vert
\mathbf{b}_{1}\Vert^{2} \Vert \mathbf{b}_{2}\Vert^{2}...\Vert
\mathbf{b}_{N_{r}}\Vert^{2}$ is minimized. Because $ \det \mathbf{B}^{\mathtt{H}} \mathbf{B}=\det
\left( \mathbf{H}^{+}\right)^{\mathtt{H}} \mathbf{H}^{+} $ is given, the best basis reduction is the
reduction which maximizes $\frac{\Vert \mathbf{b}_{1}\Vert^{2} \Vert \mathbf{b}_{2}\Vert^{2}\cdot ... \cdot \Vert
\mathbf{b}_{N_{r}}\Vert^{2}}
{\vert \det \mathbf{B}^{\mathtt{H}} \mathbf{B} \vert}$, or in other words, minimizes the
orthogonality defect.

In practice, we use discrete values for the rate, and sometimes, we should assign the rate zero to some users (when their channel is very bad). In this case, for the rate assignment for other users, we use the lattice reduction on the corresponding sublattice. It should be noted that the average transmit power is fixed per channel realization and no long-run averaging is considered, and no long-run power allocation is used.

\section{Diversity and Outage Probability}

In this section, we consider the asymptotic behavior ($\rho
\longrightarrow \infty$) of the symbol error rate (SER) for the proposed
method and the perturbation technique.  We show that for both
of these methods, the asymptotic slope of the SER
curve is equal to the number of transmit antennas. By considering
the outage probability of a fixed-rate MIMO broadcast system, we
will show that for the SER curve in high SNR, the
slope obtained by the proposed method has the largest achievable
value.
Also, we analyze the asymptotic behavior of the outage probability for the case of fixed sum-rate. We show that in this case, the slope of the corresponding curve is equal to the product of the number of transmit antennas and the number of single-antenna users.

\subsection{Fixed-rate users}

When we have the Channel-State Information (CSI) at the
transmitter, without any assumption on the transmission rates, the
outage probability is not meaningful. However, when we consider
given rates $ R_{1},...,R_{N_{r}} $ for different users, we can define
the outage probability $ P_{out} $ as the probability that the
point $ (R_{1},...,R_{N_{r}}) $ is outside the capacity region.

\begin{thm}
For a MIMO broadcast system with $ N_{t} $ transmit antennas, $ N_{r} $
single-antenna receivers ($ N_{t} \geq N_{r}$), and given rates $ R_{1},...,R_{N_{r}} $,

\begin{equation} \lim_{\rho\rightarrow \infty} \dfrac{-\log P_{out}}{\log \rho} \leq N_{t}.\end{equation}
\end{thm}

\begin{proof} Define $ P_{out1} $ as the probability that the capacity of
the point-to-point system corresponding to the first user
(consisting of $ N_{t} $ transmit antennas and one receive antenna
with independent channel coefficients and CSI at the transmitter)
is less than $ R_{1} $:

\begin{equation} P_{out1}= \Pr \lbrace \log \left( 1+\rho \Vert \mathbf{h}_{1} \Vert^{2} \right) \leq R_{1} \rbrace \end{equation}
where $ \mathbf{h}_{1}$ is the vector defined by the first row of $ \mathbf{H}$. Note that the entries of $ \mathbf{h}_{1}$ have iid complex Gaussian distribution with unit variance. Thus, its square norm has a chi square distribution. We have,
\begin{equation} \label{eq:21}       \Pr \left\lbrace \log \left( 2\rho \Vert \mathbf{h}_{1} \Vert^{2} \right) \leq R_{1} \right\rbrace \end{equation}
\begin{equation} =\Pr \left\lbrace  \Vert \mathbf{h}_{1} \Vert^{2} \leq \frac{2^{R_{1}}}{ 2 \rho} \right\rbrace \end{equation}

\begin{equation}
=  \int_0^{\frac{2^{R_{1}}}{ 2 \rho}} f_{\Vert \mathbf{h}_{1} \Vert^{2}}( x)\,dx
\end{equation}

\begin{equation}
=  \int_0^{\frac{2^{R_{1}}}{ 2 \rho}} \frac{1}{(N_{t}-1)!} x^{N_{t}-1}e^{-x}\,dx
\end{equation}
We are interested in the  large values of  $\rho$. For $\rho > 2^{R_{1}-1}$, 

\begin{equation}
\int_0^{\frac{2^{R_{1}}}{ 2 \rho}} \frac{1}{(N_{t}-1)!} x^{N_{t}-1}e^{-x}\,dx \geq \int_0^{\frac{2^{R_{1}}}{ 2 \rho}} \frac{1}{(N_{t}-1)!} x^{N_{t}-1}e^{-1}\,dx 
\end{equation}
\begin{equation}
=  \frac{e^{-1}}{(N_{t}-1)!} \int_0^{\frac{2^{R_{1}}}{ 2 \rho}} x^{N_{t}-1}\,dx
\end{equation}

\begin{equation} \label{eq:27}      =\frac{2^{N_{t}R_{1}}c}{ \rho^{N_{t}}} \end{equation}
where $c=\frac{e^{-1}}{2^{N_{t}}N_{t}!}$ is a constant number. Now,
\begin{equation}
\log\left( 1+\rho \Vert \mathbf{h}_{1} \Vert^{2} \right) \leq \log\left( 2\rho \Vert \mathbf{h}_{1} \Vert^{2} \right)  \;\;\;\; {\rm for} \;\; \rho > \frac{1}{\Vert \mathbf{h}_{1} \Vert^{2}}
\end{equation}

\begin{equation} \Longrightarrow \lim_{\rho\rightarrow \infty} \dfrac{-\log \Pr \lbrace \log\left( 1+\rho \Vert \mathbf{h}_{1} \Vert^{2} \right) \leq R_{1} \rbrace}{\log \rho} \end{equation}

\begin{equation} \leq \lim_{\rho\rightarrow \infty} \dfrac{-\log \Pr \lbrace \log\left( 2\rho \Vert \mathbf{h}_{1} \Vert^{2} \right) \leq R_{1} \rbrace}{\log \rho}\end{equation}

\begin{equation} \leq \lim_{\rho\rightarrow \infty}  \dfrac{-\log \frac{2^{N_{t}R_{1}}c}{ \rho^{N_{t}}}}{\log \rho} =N_{t}
\end{equation}

\begin{equation} \Longrightarrow \lim_{\rho\rightarrow \infty} \dfrac{-\log P_{out1}}{\log \rho} \leq N_{t}. \end{equation}

According to the definition of $P_{out1} $, $P_{out} \geq P_{out1}$. Therefore,
\begin{equation}  \lim_{\rho\rightarrow \infty} \dfrac{-\log P_{out}}{\log \rho} \leq  \lim_{\rho\rightarrow \infty} \dfrac{-\log P_{out1}}{\log \rho} \leq N_{t}. \end{equation}

\end{proof}

We can define the diversity gain of a MIMO broadcast
constellation or its \textit{precoding diversity} as $
\lim_{\rho\rightarrow \infty} \dfrac{- \log P_{e}}{\log \rho}$ where
$ P_{e} $ is the probability of error. Similar to \cite [lemma 5]{ZT2003}, we can bound the precoding diversity by $ \lim_{\rho\rightarrow \infty} \dfrac{-\log P_{out}}{\log \rho} $ . Thus, based on theorem 2, the
maximum achievable diversity is $ N_{t} $. 

We  show that the proposed
method (based on lattice-basis reduction) achieves the maximum
precoding diversity. 
To prove this, in lemma 1 and lemma 2, we relate the length of the largest vector of the  reduced basis $ \mathbf{B}$ to $ d_{\mathbf{H}^{\mathtt{H}}}$ (the minimum distance of the lattice generated by $ \mathbf{H}^{\mathtt{H}}$). In lemma 3, we bound the probability that $ d_{\mathbf{H}^{\mathtt{H}}}$ is too small. Finally, in theorem 3, we prove the main result by relating the minimum distance of the receive constellation to the length of the largest vector of the  reduced basis $ \mathbf{B}$, and combining the bounds on the probability that $ d_{\mathbf{H}^{\mathtt{H}}}$ is too small, and the probability that the noise vector is too large.

\begin{lemma}
Consider $ \mathbf{B}=[\mathbf{b}_{1}...\mathbf{b}_{M}] $ as an $ N
\times M $ matrix, with the orthogonality defect $ \delta $,  and $
\mathbf{B}^{-\mathtt{H} }= [\mathbf{a}_{1}...\mathbf{a}_{M}]$ as the
inverse of its Hermitian (or its pseudo-inverse if $M<N$). Then,
\begin{equation} \label{eq:lemma1-1}
\max\lbrace \Vert \mathbf{b}_{1}\Vert,...,\Vert \mathbf{b}_{M}\Vert \rbrace \leq \frac{\sqrt{\delta}}{\min\lbrace \Vert \mathbf{a}_{1}\Vert,...,\Vert \mathbf{a}_{M}\Vert \rbrace} 
\end{equation}
and
\begin{equation} \label{eq:lemma1-2}
\max\lbrace \Vert \mathbf{a}_{1}\Vert,...,\Vert \mathbf{a}_{M}\Vert \rbrace \leq \frac{\sqrt{\delta}}{\min\lbrace \Vert \mathbf{b}_{1}\Vert,...,\Vert \mathbf{b}_{M}\Vert \rbrace} .
\end{equation}
\end{lemma}

\begin{proof} Consider $ \mathbf{b}_{i} $ as an arbitrary column of $
\mathbf{B} $. The vector $ \mathbf{b}_{i} $ can be written as $
\mathbf{b}_{i}^{\prime} + \sum_{i \neq j} c_{i,j} \mathbf{b}_{j} $,
where $ \mathbf{b}_{i}^{\prime} $ is orthogonal to $ \mathbf{b}_{j}
$ for $ i \neq j $. Now, $ [\mathbf{b}_{1}...\mathbf{b}_{i-1} \mathbf{b}_{i}^{\prime} \mathbf{b}_{i+1}...\mathbf{b}_{M}]$ can be written as $ \mathbf{BP}$ where $ \mathbf{P}$ is a unit-determinant $ M \times M$ matrix (a column operation matrix):
\begin{equation} \Vert \mathbf{b}_{1} \Vert^{2}...\Vert\mathbf{b}_{i-1}\Vert^{2} . \Vert \mathbf{b}_{i} \Vert^{2} . \Vert \mathbf{b}_{i+1}\Vert^{2}...\Vert \mathbf{b}_{M} \Vert^{2} \end{equation}
\begin{equation}
= \delta \det \mathbf{B^{\mathtt{H} }B} = \delta  \det \mathbf{P}^{\mathtt{H} }\mathbf{B^{\mathtt{H} }}  \mathbf{B} \mathbf{P} 
\end{equation}
 \begin{equation} = \delta \det\left(  [\mathbf{b}_{1}...\mathbf{b}_{i-1} \mathbf{b}_{i}^{\prime} \mathbf{b}_{i+1}...\mathbf{b}_{M}]^{\mathtt{H} }[\mathbf{b}_{1}...\mathbf{b}_{i-1} \mathbf{b}_{i}^{\prime} \mathbf{b}_{i+1}...\mathbf{b}_{M}]\right) .  \end{equation}
According to the Hadamard theorem:
\begin{equation}
\det\left(  [\mathbf{b}_{1}...\mathbf{b}_{i-1} \mathbf{b}_{i}^{\prime} \mathbf{b}_{i+1}...\mathbf{b}_{M}]^{\mathtt{H} }[\mathbf{b}_{1}...\mathbf{b}_{i-1} \mathbf{b}_{i}^{\prime} \mathbf{b}_{i+1}...\mathbf{b}_{M}]\right)  \leq
 \end{equation}
 \begin{equation}
 \Vert \mathbf{b}_{1} \Vert^{2}...\Vert\mathbf{b}_{i-1}\Vert^{2} . \Vert \mathbf{b}_{i}^{\prime} \Vert^{2} . \Vert \mathbf{b}_{i+1}\Vert^{2}...\Vert \mathbf{b}_{M} \Vert^{2}.
\end{equation}
Therefore,
\begin{equation}
\Vert \mathbf{b}_{1} \Vert^{2}...\Vert\mathbf{b}_{i-1}\Vert^{2} . \Vert \mathbf{b}_{i} \Vert^{2} . \Vert \mathbf{b}_{i+1}\Vert^{2}...\Vert \mathbf{b}_{M} \Vert^{2}\leq   \delta \Vert \mathbf{b}_{1} \Vert^{2}...\Vert\mathbf{b}_{i-1}\Vert^{2} . \Vert \mathbf{b}_{i}^{\prime} \Vert^{2} . \Vert \mathbf{b}_{i+1}\Vert^{2}...\Vert \mathbf{b}_{M} \Vert^{2}  
\end{equation}
\begin{equation}
\Longrightarrow  \Vert \mathbf{b}_{i} \Vert \leq  \sqrt{\delta} \Vert \mathbf{b}_{i}^{\prime} \Vert.
\end{equation}
Also, $ \mathbf{B}^{+}\mathbf{B}=\mathbf{I}$ results in
$<\!\!\mathbf{a}_{i},\mathbf{b}_{i}\!\!>\;= 1$ and
$<\!\!\mathbf{a}_{i},\mathbf{b}_{j}\!\!>\;= 0 $ for $ i\neq j $.
Therefore,
\begin{equation}  1 \; =\;\;<\!\!\mathbf{a}_{i},\mathbf{b}_{i}\!\!>\;\;= \;\; <\!\!\mathbf{a}_{i},(\mathbf{b}_{i}^{\prime} + \sum_{i \neq j} c_{i,j} \mathbf{b}_{j})\!\!> \;\;= \;\; <\!\!\mathbf{a}_{i},\mathbf{b}_{i}^{\prime}\!\!> \;\; 
\end{equation}
Now, $\mathbf{a}_{i}$ and $\mathbf{b}_{i}^{\prime} $, both are orthogonal to the $(M-1)$-dimensional subspace generated by the vectors $\mathbf{b}_{j} $ ($ j\neq i$). Thus, 
\begin{equation}
1 = \;\; <\!\!\mathbf{a}_{i},\mathbf{b}_{i}^{\prime}\!\!> \;\; \; = \; \Vert\mathbf{a}_{i}\Vert .\Vert\mathbf{b}_{i}^{\prime}\Vert \geq \Vert\mathbf{a}_{i}\Vert . \frac{\Vert\mathbf{b}_{i}\Vert}{\sqrt{\delta}}\end{equation}
\begin{equation} \Longrightarrow 1 \geq \Vert\mathbf{b}_{i}\Vert . \frac{\Vert\mathbf{a}_{i}\Vert}{\sqrt{\delta}} \end{equation}
\begin{equation} \label{eq:15}  \Longrightarrow \Vert\mathbf{b}_{i}\Vert \leq \frac{\sqrt{\delta}}{\Vert\mathbf{a}_{i}\Vert} \end{equation}
The above relation is valid for every $ i $, $ 1\leq i\leq M $.
Without loss of generality, we can assume that $
\max\lbrace \Vert \mathbf{b}_{1}\Vert,...,\Vert \mathbf{b}_{M}\Vert
\rbrace= \Vert \mathbf{b}_{k} \Vert $:
\begin{equation} \max\lbrace \Vert \mathbf{b}_{1}\Vert,...,\Vert \mathbf{b}_{M}\Vert \rbrace= \Vert \mathbf{b}_{k} \Vert \leq \frac{\sqrt{\delta}}{\Vert\mathbf{a}_{k}\Vert} \end{equation}
  \begin{equation}\leq \frac{\sqrt{\delta}}{\min\lbrace \Vert \mathbf{a}_{1}\Vert,...,\Vert \mathbf{a}_{M}\Vert \rbrace} . \end{equation}
    Similarly, by using (\ref{eq:15}), we can also obtain the following inequality:
 \begin{equation} 
\max\lbrace \Vert \mathbf{a}_{1}\Vert,...,\Vert \mathbf{a}_{M}\Vert \rbrace \leq \frac{\sqrt{\delta}}{\min\lbrace \Vert \mathbf{b}_{1}\Vert,...,\Vert \mathbf{b}_{M}\Vert \rbrace} .
\end{equation} 
\end{proof}

\begin{lemma}
Consider $ \mathbf{B}=[\mathbf{b}_{1}...\mathbf{b}_{M}] $ as an
LLL-reduced basis for the lattice generated by $ \mathbf{H}^{+} $
and $ d_{\mathbf{H}^{\mathtt{H}}} $ as the minimum distance of the lattice
generated by $ \mathbf{H}^{\mathtt{H}} $. Then, there is a constant $ \alpha_{M} $
(independent of $ \mathbf{H} $) such that
\begin{equation} \max\lbrace \Vert \mathbf{b}_{1}\Vert,...,\Vert \mathbf{b}_{M}\Vert \rbrace \leq \frac{\alpha_{M}}{d_{\mathbf{H}^{\mathtt{H}}}} . \end{equation}
\end{lemma}

\begin{proof} According to
the theorem 1,
\begin{equation} \label{eq:delta} \sqrt{\delta} \leq 2^{M(M-1)}. \end{equation}
Consider $ \mathbf{B}^{-\mathtt{H}}=[\mathbf{a}_{1},...,
\mathbf{a}_{M}] $. By using lemma 1 and (\ref{eq:delta}),
\begin{equation}\label{eq:36}
\max\lbrace \Vert \mathbf{b}_{1}\Vert,...,\Vert \mathbf{b}_{M}\Vert \rbrace \leq \frac{\sqrt{\delta}}{\min\lbrace \Vert \mathbf{a}_{1}\Vert,...,\Vert \mathbf{a}_{M}\Vert \rbrace} \leq \frac{2^{M(M-1)}}{\min\lbrace \Vert \mathbf{a}_{1}\Vert,...,\Vert \mathbf{a}_{M}\Vert \rbrace}
\end{equation}
The basis $ \mathbf{B}$ can be written as $ \mathbf{B}=\mathbf{H}^{+}\mathbf{U}$ for some unimodular matrix $\mathbf{U}$:
\begin{equation}
\mathbf{B}^{-\mathtt{H}} = ((\mathbf{H}^{+}\mathbf{U})^{\mathtt{H}})^{+}=(\mathbf{U}^{\mathtt{H}}\mathbf{H}^{-\mathtt{H}})^{+}=\mathbf{H}^{\mathtt{H}}\mathbf{U}^{-\mathtt{H}}.
\end{equation}
Noting that $\mathbf{U}^{-\mathtt{H}} $ is unimodular, $ \mathbf{B}^{-\mathtt{H}}=[\mathbf{a}_{1},...,
\mathbf{a}_{M}] $ is another basis for the lattice generated by $\mathbf{H}^{\mathtt{H}}$. Therefore, the vectors $\mathbf{a}_{1},...,
\mathbf{a}_{M}$ are vectors from the lattice generated by $\mathbf{H}^{\mathtt{H}}$, and therefore, the length of each of them is at least $d_{\mathbf{H}^{\mathtt{H}}}$:
\begin{equation}
\Vert \mathbf{a}_{i} \Vert \geq d_{\mathbf{H}^{\mathtt{H}}}  \; \; \;\; \mathrm{for} \; \; 1\leq i \leq M
\end{equation}
\begin{equation} \label{eq:39}
 \Longrightarrow \min\lbrace \Vert \mathbf{a}_{1}\Vert,...,\Vert \mathbf{a}_{M}\Vert \rbrace \geq d_{\mathbf{H}^{\mathtt{H}}}
\end{equation}

\begin{equation}
\mathrm{(\ref{eq:36}) \; and \; (\ref{eq:39})} \Longrightarrow \max\lbrace \Vert \mathbf{b}_{1}\Vert,...,\Vert \mathbf{b}_{M}\Vert \rbrace \leq  \frac{2^{M(M-1)}}{d_{\mathbf{H}^{\mathtt{H}}}}.
\end{equation}
\end{proof}

\begin{lemma}
Assume that the entries of the $ N\times M $ matrix $\mathbf{H}$ has
independent complex Gaussian distribution with zero mean and unit
variance and consider $ d_{\mathbf{H}} $ as the minimum distance of the
lattice generated by $ \mathbf{H} $. Then, there is a constant $\beta_{N,M} $
such that

\begin{eqnarray}
\Pr\left\lbrace d_{\mathbf{H}}\leq \varepsilon \right\rbrace \leq \left \lbrace
\begin{array}{c}
 \beta_{N,M} \varepsilon^{2N}   \;\;\;\;\;\;\;\;\;\;\;\;\;\;\;\;\;\;\;\;\;\;\;\;\;\;\;\;\;\;\;\;\;\; \textmd{for}  \; M<N  \\
\beta_{N,N} \varepsilon^{2N}.\max\left\lbrace -(\ln \varepsilon)^{N+1},1\right\rbrace \;\; \textmd{for}
\;M=N
\end{array} \right. .
\end{eqnarray}

 \end{lemma}

Proof: See Appendix B.

\begin{thm}
For a MIMO broadcast system with $ N_{t} $ transmit antennas and $ N_{r} $
single-antenna receivers ($ N_{t} \geq N_{r}$) and fixed rates $ R_{1},...,R_{N_{r}} $, using
the lattice-basis-reduction method,
\begin{equation} \lim_{\rho\rightarrow \infty} \dfrac{-\log P_{e}}{\log \rho}=N_{t}.\end{equation}
\end{thm}

\begin{proof} Consider $ \mathbf{B}=[\mathbf{b}_{1}...\mathbf{b}_{N_{r}}] $ as
the LLL-reduced basis for the lattice generated by
$\mathbf{H}^{+} $. Each transmitted vector $\mathbf{s}$ is inside
the parallelotope, generated by $
r_{1}\mathbf{b}_{1},...,r_{N_{r}}\mathbf{b}_{N_{r}}$ (where $
r_{1},...,r_{N_{r}} $ are constant values determined by the rates of
the users). Thus, every transmitted vector $\mathbf{s}$ can be
written as
\begin{equation}\mathbf{s} = t_{1}\mathbf{b}_{1}+...+t_{N_{r}}\mathbf{b}_{N_{r}}, \;\; \; \frac{-r_{i}}{2}\leq t_{i}\leq \frac{r_{i}}{2}.\end{equation}
For each of the transmitted vectors, the energy is
\begin{equation} P =\Vert\mathbf{s}\Vert^{2} =\Vert t_{1}\mathbf{b}_{1}+...+t_{N_{r}}\mathbf{b}_{N_{r}}\Vert^{2}\end{equation}
\begin{equation}
\Longrightarrow P \leq \left( \Vert t_{1}\mathbf{b}_{1}\Vert +...+\Vert t_{N_{r}}\mathbf{b}_{N_{r}}\Vert\right) ^{2}
\end{equation}
\begin{equation} \Longrightarrow P  \leq \left( \frac{r_{1}}{2} \Vert\mathbf{b}_{1}\Vert +...+\frac{r_{N_{r}}}{2}\Vert\mathbf{b}_{N_{r}}\Vert\right) ^{2}.\end{equation}
 Thus, the average transmitted energy is
 \begin{equation} P_{av} = {\rm E}(P)\leq N_{r}^{2} \left(\max \left\lbrace \frac{r_{1}}{2} \Vert \mathbf{b}_{1}\Vert,...,\frac{r_{N_{r}}}{2}\Vert \mathbf{b}_{N_{r}}\Vert \right\rbrace \right)^{2}  \leq   c_{1}. (\max\lbrace \Vert \mathbf{b}_{1}\Vert^{2},...,\Vert \mathbf{b}_{N_{r}}\Vert^{2} \rbrace)\end{equation}
where $ c_{1}= \frac{N_{r}^{2}}{4}\max \left\lbrace r_{1}^{2},...,r_{N_{r}}^{2}\right\rbrace $. The
received signals (without the effect of noise) are points from the
$\mathbb{Z}^{2N_{r}}$ lattice. If we consider the normalized system (by scaling the signals such that the average transmitted energy becomes equal to one),
\begin{equation} \label{eq:60}
d^{2} = \dfrac{1}{P_{av}}\geq
\dfrac{1}{c_{1}. (\max\lbrace \Vert \mathbf{b}_{1}\Vert^{2},...,\Vert \mathbf{b}_{N_{r}}\Vert^{2} \rbrace)}\end{equation}
 is the squared distance between
the received signal points.

For the normalized system, $ \dfrac{1}{\rho} $ is the energy of
the noise at each receiver and $ \dfrac{1}{2 \rho} $ is the energy of the noise per each real dimension. Using (\ref{eq:60}), for any positive number $ \gamma$,
$$ \Pr \left\lbrace d^{2}\leq \dfrac{\gamma }{\rho} \right\rbrace  $$
\begin{equation} \label{eq:51} \leq \Pr \left\lbrace \dfrac{1}{c_{1} \max\lbrace \Vert \mathbf{b}_{1}\Vert^{2},...,\Vert \mathbf{b}_{N_{r}}\Vert^{2} \rbrace}\leq \dfrac{\gamma }{\rho} \right\rbrace \end{equation}\\
Using lemma 2,
\begin{equation} \label{eq:52} \max\lbrace \Vert \mathbf{b}_{1}\Vert,...,\Vert \mathbf{b}_{N_{r}}\Vert \rbrace \leq \frac{\alpha_{N_{r}}}{d_{\mathbf{H}^{\mathtt{H}}}} \end{equation}

\begin{equation}(\ref{eq:51}), (\ref{eq:52}) \Longrightarrow \Pr \left\lbrace d^{2}\leq \dfrac{\gamma }{\rho} \right\rbrace \leq \Pr\left\lbrace \dfrac{d_{\mathbf{H}^{\mathtt{H}}}^{2}}{c_{1}\alpha_{N_{r}}^{2}}\leq \dfrac{\gamma }{\rho} \right\rbrace  =  \Pr \left\lbrace d_{\mathbf{H}^{\mathtt{H}}}^{2}\leq \dfrac{\gamma c_{1}\alpha_{N_{r}}^{2}}{\rho} \right\rbrace \end{equation}\\

The $N_{t}\times N_{r} $ matrix $\mathbf{H}^{\mathtt{H}} $ has independent complex Gaussian distribution with zero mean and unit
variance. Therefore, by using lemma 3 (considering $M=N_{r}$ and $ N=N_{t}$), we can bound the probability that $ d_{\mathbf{H}^{\mathtt{H}}}$ is too small.

\textbf{Case 1}, $N_{r}=N_{t}$:

\begin{equation} \Pr\left\lbrace d^{2}\leq \dfrac{\gamma }{\rho} \right\rbrace  \leq  \Pr\left\lbrace d_{\mathbf{H}^{\mathtt{H}}}^{2}\leq \dfrac{\gamma c_{1}\alpha_{N_{t}}^{2}}{\rho} \right\rbrace\end{equation}

\begin{equation}
\leq \beta_{N_{t},N_{t}} \left( \frac{\gamma c_{1}\alpha_{N_{t}}^{2}}{\rho}\right) ^{N_{t}} \max\left\lbrace \left( -\frac{1}{2}\ln \frac{\gamma c_{1}\alpha_{N_{t}}^{2}}{\rho}\right)^{N_{t}+1},1\right\rbrace 
\end{equation}

\begin{equation}  \leq \beta_{N_{t},N_{t}} \left( \frac{\gamma c_{1}\alpha_{N_{t}}^{2}}{\rho}\right) ^{N_{t}} \max \left\lbrace \left(\ln \rho\right)^{N_{t}+1},1\right\rbrace   \;\;\;{\rm for} \;  \gamma>1 \;{\rm and} \;\rho>  \frac{1}{c_{1}\alpha_{N_{t}}^{2}} \end{equation}
\begin{equation} \label{eq:55} \leq   \frac{c_{2}\gamma^{N_{t}}}{\rho ^{N_{t}}}  \left(\ln \rho\right)^{N_{t}+1}   \;\;\;{\rm for} \;  \gamma>1 \;{\rm and} \;\rho>  \max\left\lbrace \frac{1}{c_{1}\alpha_{N_{t}}^{2}},e\right\rbrace  \end{equation}
where $  c_{2} $ is a constant number and $e$ is the Euler number.

If the magnitude of the noise component in each real dimension is less
than $ \frac{1}{2}d $, the transmitted data will be decoded
correctly. Thus, we can bound the probability of error by the
probability that $ \vert w_{i} \vert^{2} $ is greater than $
\frac{1}{4}d^{2}$ for at least one $ i$, $ 1 \leq i \leq 2N_{t} $.
Therefore, using the union bound,

\begin{equation}    P_{e}\leq   2N_{t} \left(  \Pr\left\lbrace \vert w_{1} \vert^{2} \geq \frac{1}{4}d^{2}  \right\rbrace \right)  \end{equation}

$$ = 2N_{t} \left(  \Pr\left\lbrace d^{2}\leq \dfrac{4}{\rho} \right\rbrace . \Pr\left\lbrace \vert w_{1} \vert^{2} \geq \frac{1}{4}d^{2} \left\vert d^{2}\leq \dfrac{4}{\rho} \right. \right\rbrace \right.$$
$$+ \Pr\left\lbrace \dfrac{4}{\rho} \leq d^{2}\leq \dfrac{8}{\rho} \right\rbrace . \Pr\left\lbrace \vert w_{1} \vert^{2} \geq \frac{1}{4}d^{2} \left\vert \dfrac{4}{\rho} \leq d^{2}\leq \dfrac{8}{\rho}\right. \right\rbrace
$$
\begin{equation}\left.  + \Pr\left\lbrace \dfrac{8}{\rho} \leq d^{2}\leq \dfrac{16}{\rho}  \right\rbrace . \Pr\left\lbrace \vert w_{1} \vert^{2} \geq \frac{1}{4}d^{2} \left\vert \dfrac{8}{\rho} \leq d^{2}\leq \dfrac{16}{\rho}\right. \right\rbrace + ... \right)\end{equation}

 $$\leq 2N_{t} \left(  \Pr\left\lbrace d^{2}\leq \dfrac{4}{\rho} \right\rbrace + \Pr\left\lbrace \dfrac{4}{\rho} \leq d^{2}\leq \dfrac{8}{\rho} \right\rbrace . \Pr\left\lbrace \vert w_{1} \vert^{2} \geq \frac{1}{4}.\frac{4}{\rho}  \right\rbrace
\right. $$
 \begin{equation}\left.  + \Pr\left\lbrace \dfrac{8}{\rho} \leq d^{2}\leq \dfrac{16}{\rho}  \right\rbrace . \Pr\left\lbrace \vert w_{1} \vert^{2} \geq \frac{1}{4}.\frac{8}{\rho}  \right\rbrace + ... \right)\end{equation}

$$\leq 2N_{t} \left(  \Pr\left\lbrace d^{2}\leq \dfrac{4}{\rho}  \right\rbrace
+ \Pr\left\lbrace d^{2}\leq \dfrac{8}{\rho}  \right\rbrace .
\Pr\left\lbrace \vert w_{1} \vert^{2} \geq \frac{1}{\rho}
\right\rbrace \right. $$
 \begin{equation} \label{eq:56}
 \left.  + \Pr\left\lbrace d^{2}\leq \dfrac{16}{\rho}  \right\rbrace . \Pr\left\lbrace \vert w_{1} \vert^{2} \geq \frac{2}{\rho}  \right\rbrace + ... \right)\end{equation}

 For the product terms in (\ref{eq:56}), we can bound the first part by (\ref{eq:55}). To bound the second part, we note that $ w_{1}$ has real Gaussian distribution with variance $\frac{1}{2\rho} $. Therefore,

\begin{equation}\label{eq:59} \Pr\left\lbrace \vert w_{1} \vert^{2} \geq \frac{\theta}{\rho}  \right\rbrace = Q(\sqrt{2\theta}) \leq e^{-\theta}\end{equation}
Now, for $\rho>  \max\left\lbrace \frac{1}{c_{1}\alpha_{N_{t}}^{2}},e\right\rbrace$,
\begin{equation} (\ref{eq:55}),\;(\ref{eq:56})\;{\rm and} \;(\ref{eq:59})\Longrightarrow  P_{e}\leq  2N_{t}\left(  \Pr\left\lbrace \vert w_{1} \vert^{2} \geq \frac{1}{4}d^{2} \right\rbrace  \right)\end{equation}

  \begin{equation} \leq 2N_{t} \left(  \dfrac{4^{N_{t}}c_{2}}{\rho^{N_{t}}} \left(\ln \rho \right)^{N_{t}+1} + \sum_{i=0}^{\infty} \dfrac{2^{N_{t}(i+3)}c_{2}}{\rho^{N_{t}}} \left(\ln \rho \right)^{N_{t}+1} e^{-2^{i}}  \right)    \end{equation}
 
\begin{equation} \leq \frac{(\ln \rho)^{N_{t}+1}}{\rho^{N_{t}}} . c_{2}. 2N_{t} \left(4^{N_{t}}+ \sum_{i=0}^{\infty}2^{N_{t}(i+3)} e^{-2^{i}}  \right)     \end{equation}

\begin{equation}\leq  \dfrac{c_{3}(\ln \rho)^{N_{t}+1}}{\rho^{N_{t}}}  \end{equation}
where $ c_{3} $ is a constant number which only depends on
$N_{t} $. Thus,

\begin{equation} \lim_{\rho\rightarrow \infty} \dfrac{-\log P_{e}}{\log \rho} \geq \lim_{\rho\rightarrow \infty} \dfrac{N_{t}\log \rho -\log (\ln \rho)^{N_{t}+1}-\log c_{3}}{\log \rho} = N_{t}.  \end{equation}\\
According to Theorem 2, this limit can not be greater than $ N_{t} $.
Therefore,

\begin{equation} \lim_{\rho\rightarrow \infty} \dfrac{-\log P_{e}}{\log \rho} =N_{t}.  \end{equation}

\textbf{Case 2}, $N_{r}<N_{t}$: 

For the $N_{t}\times N_{r} $ matrix $\mathbf{H}^{\mathtt{H}} $, we use the first inequality in lemma 3 (by considering $M=N_{r}$ and $N=N_{t}$) to bound the probability that $ d_{\mathbf{H}^{\mathtt{H}}}$ is too small:

\begin{equation} \Pr\left\lbrace d^{2}\leq \dfrac{\gamma }{\rho} \right\rbrace  \leq  \Pr\left\lbrace d_{\mathbf{H}^{\mathtt{H}}}^{2}\leq \dfrac{\gamma c_{1}\alpha_{N_{r}}^{2}}{\rho} \right\rbrace\end{equation}

\begin{equation}
\leq \beta_{N_{t},N_{r}} \left( \frac{\gamma c_{1}\alpha_{N_{r}}^{2}}{\rho}\right) ^{N_{t}} 
\end{equation}

\begin{equation}  \leq \beta_{N_{t},N_{r}} \left( \frac{\gamma c_{1}\alpha_{N_{r}}^{2}}{\rho}\right) ^{N_{t}}    \;\;\;{\rm for} \;  \gamma>1 \;{\rm and} \;\rho>  \frac{1}{c_{1}\alpha_{N_{r}}^{2}} \end{equation}
\begin{equation}  \leq   \frac{c_{2}\gamma^{N_{t}}}{\rho ^{N_{t}}}   \;\;\;{\rm for} \;  \gamma>1 \;{\rm and} \;\rho>  \frac{1}{c_{1}\alpha_{N_{r}}^{2}}  \end{equation}.

The rest of proof is similar to the case 1.

\end{proof}

\begin{Corollary}
Perturbation technique achieves the maximum precoding diversity in
fixed-rate MIMO broadcast systems.
\end{Corollary}
\begin{proof} In the perturbation technique, for the transmission of each
data vector $\mathbf{u}$, among the set $\left\lbrace
\mathbf{H}^{+}(\mathbf{u} + a \mathbf{l})\Vert \mathbf{l}\in
\mathbb{Z}^{2N_{r}} \right\rbrace $, the nearest point to the origin
is chosen. The transmitted vector in the lattice-reduction-based
method belongs to that set. Therefore, the energy of the
transmitted signal in the lattice-reduction-based method can not
be less than the transmitted energy in the perturbation technique.
Thus, the average transmitted energy for the perturbation method
is at most equal to the average transmitted energy of the
lattice-reduction-based method. The rest of the proof is the same
as the proof of theorem 3.
\end{proof}

\subsection{Fixed sum-rate}

When the sum-rate $ R_{sum}$ is given, similar to the previous part, we can define the outage probability as the probability that the sum-capacity of the broadcast system is less than $ R_{sum}$. 

\begin{thm}
For a MIMO broadcast system with $ N_{t} $ transmit antennas, $ N_{r} $
single-antenna receivers, and a given sum-rate $ R_{sum} $,

\begin{equation} \lim_{\rho\rightarrow \infty} \dfrac{-\log P_{out}}{\log \rho} \leq N_{t}N_{r}.\end{equation}
\end{thm}

\begin{proof}

For any channel matrix $\mathbf{H}$, we have \cite{VJG}

\begin{equation} C_{sum}=  \sup_\mathbf{D}  \log \vert  \mathbf{I}_{N_{r}}+\rho \mathbf{H}^{\mathtt{H}}\mathbf{D}\mathbf{H} \vert  \end{equation}
where $ \mathbf{D}$ is a diagonal matrix with non-negative elements and unit trace. Also, \cite{Matrix}

\begin{equation} \label{eq:74} \vert  2\rho \mathbf{H}^{\mathtt{H}}\mathbf{D}\mathbf{H} \vert \leq \frac{  \left( 2\rho{\rm tr}  \mathbf{H}^{\mathtt{H}}\mathbf{D}\mathbf{H} \right)^{N_{r}}}{N_{r}^{N_{r}}} = \frac{  \left( 2\rho{\rm tr}  \mathbf{H}^{\mathtt{H}}\mathbf{H} \right) ^{N_{r}}}{N_{r}^{N_{r}}}. \end{equation}

The entries of $ \mathbf{H}$ have iid complex Gaussian distribution with unit variance. Thus ${\rm tr} \rho \mathbf{H}^{\mathtt{H}}\mathbf{H}$ is equal to the square norm of an $N_{t}N_{r}$-dimensional complex Gaussian vector and has a chi square distribution with $2N_{t}N_{r}$ degrees of freedom. Thus, we have (similar to the equations \ref{eq:21}-\ref{eq:27}, in the proof of theorem 2),
\begin{equation} \Pr \left\lbrace \log \frac{\left(2\rho {\rm tr}  \mathbf{H}^{\mathtt{H}}\mathbf{H} \right) ^{N_{r}}}{N_{r}^{N_{r}}} \leq R_{sum} \right\rbrace \end{equation}

\begin{equation}
= \Pr \left\lbrace  {\rm tr} \mathbf{H}^{\mathtt{H}}\mathbf{H}  \leq \frac{2^{\frac{R_{sum}}{N_{r}}}N_{r}}{2 \rho} \right\rbrace
\end{equation}

\begin{equation} \label{eq:76} \geq \frac{2^{N_{t}R_{sum}}N_{r}^{N_{t}N_{r}}c}{ \rho^{N_{t}N_{r}}}  \;\;\; \;\;({\rm for} \;\rho > \frac{2^{\frac{R_{sum}}{N_{r}}}N_{r}}{2}) \end{equation}
where $c$ is a constant number. Now,
\begin{equation} \lim_{\rho\rightarrow \infty} \frac{-\log P_{out}}{\log \rho} = \end{equation}

\begin{equation} \lim_{\rho\rightarrow \infty} \frac{-\log \Pr \left\lbrace \sup_\mathbf{D}  \log \vert  \mathbf{I}_{N_{r}}+\rho \mathbf{H}^{\mathtt{H}}\mathbf{D}\mathbf{H} \vert \leq R_{sum} \right\rbrace}{\log \rho} \end{equation}

\begin{equation} \label{eq:79} \leq \lim_{\rho\rightarrow \infty} \frac{-\log \Pr \lbrace \sup_{\mathbf{D}}  \log \vert  2 \rho \mathbf{H}^{\mathtt{H}}\mathbf{D}\mathbf{H} \vert \leq R_{sum} \rbrace}{\log \rho}.\end{equation}
By using (\ref{eq:74}), (\ref{eq:76}), and (\ref{eq:79}):
\begin{equation} \lim_{\rho\rightarrow \infty} \frac{-\log P_{out}}{\log \rho} \leq \lim_{\rho\rightarrow \infty}  \dfrac{-\log \frac{2^{N_{t} R_{sum}}N_{r}^{N_{t}N_{r}}c}{ \rho^{N_{t}N_{r}}}}{\log \rho} =N_{t}N_{r}.
\end{equation}

\end{proof}

The slope $ N_{t}N_{r}$ for the SER curve can be easily achieved by sending to only the best user. Similar to the proof of theorem 3, the slope of the symbol-error rate curve is asymptoticly determined by the slope of the probability that $\vert h_{max}\vert$ is smaller than a constant number, where $ h_{max}$ is the entry of $\mathbf{H}$ with maximum norm. Due to the iid complex Gaussian distribution of the entries of $\mathbf{H}$, this probability decays with the same rate as $ \rho^{-N_{t}N_{r}}$, for large $ \rho$. However, although sending to only the best user achieves the optimum slope for the SER curve, it is not an efficient transmission technique because it reduces the capacity to the order of $\log \rho$ (instead of $N_{r} \log \rho$).

\section{Simulation Results}

Figure 2 presents the simulation results for the performance of the
proposed schemes, the perturbation scheme \cite{PeelHoch}, and the
naive channel inversion approach. The number of the transmit
antennas is $N_{t}=4$ and there are $N_{r}=4$ single-antenna users in the
system. The overall transmission rate is 8 bits per channel use,
where 2 bits are assigned to each user, i.e. a QPSK constellation is assigned to each user.

By considering the slope of the curves in figure 2, we see that by
using the proposed reduction-based schemes, we can achieve the
maximum precoding diversity, with a low complexity. Also, as
compared to the perturbation scheme, we have a negligible loss in
the performance (about 0.2 dB). Moreover, compared to the
approximated perturbation method \cite{fischer}, we have about 1.5
dB improvement by sending the bits, corresponding to the shift
vector, at the beginning of the transmission. Without sending the
shift vector, the performance of the proposed method is the same
as that of the approximated perturbation method \cite{fischer}.
The modified perturbation method (with sending two shift bits for
each user) has around 0.3 dB improvement compared to the
perturbation method.

\begin{figure}
  \centering
  \includegraphics[scale=.6,clip]{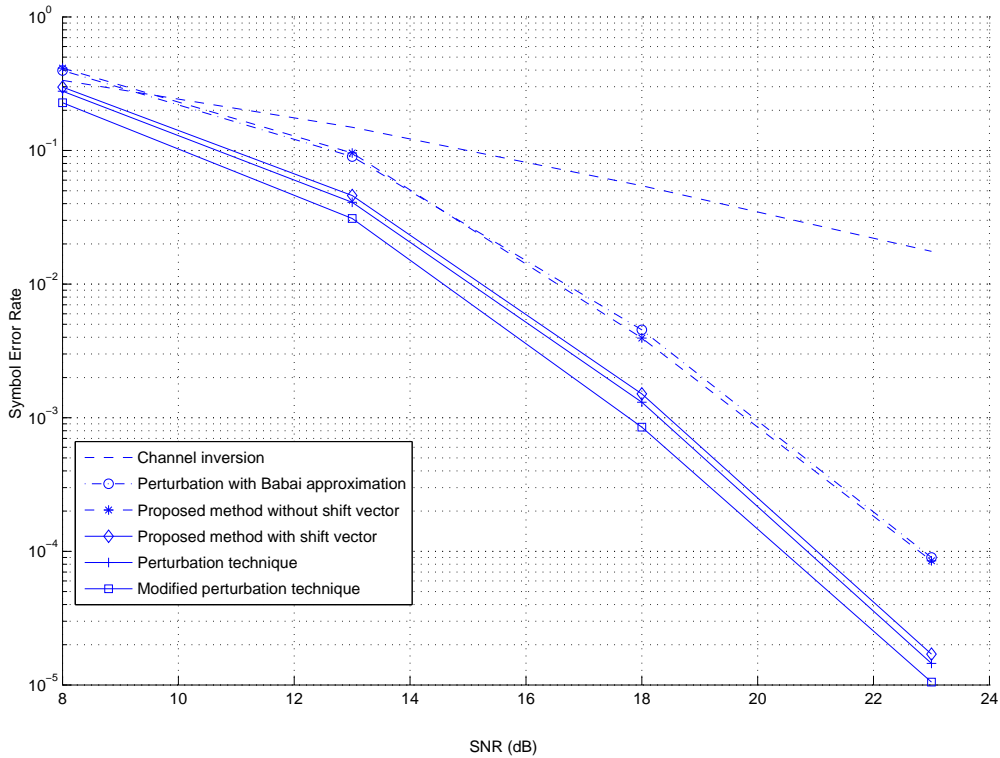}

  \caption{Symbol Error Rate of the proposed schemes, the perturbation scheme \cite{PeelHoch}, and the naive channel inversion approach for $N_{t}=4$ transmit antennas and $N_{r}=4$ single-antenna receivers with the rate $R=2$ bits per channel use per user.}
  \label{fig:reduction}
\end{figure}

Figure 3 compares the regularized proposed scheme with V-BLAST modifications of Zero-Forcing and Babai
approximation for the same setting. As shown in the simulation results (and also in the simulation results in \cite{PeelHoch} and \cite{fischer}), the modulo-MMSE-VBLAST scheme does not achieve a precoding diversity better than zero forcing (though it has a good performance in the low SNR region). However, combining the lattice-reduction-aided (LRA) scheme with MMSE-VBLAST precoding or other schemes such as regularization improves its performance by a finite coding gain (without changing the slope of the curve of symbol-error-rate). Combining both the regularization and the shift vector can result in better performance compared to other alternatives.

\begin{figure}
  \centering
  \includegraphics[scale=.6,clip]{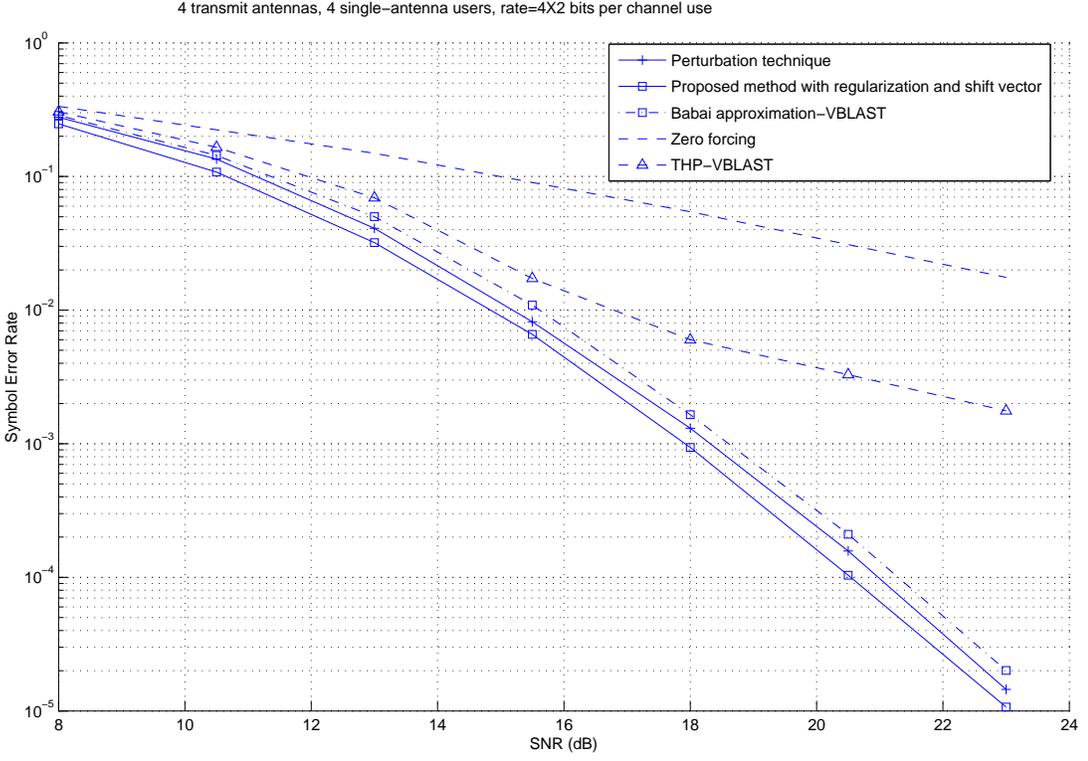}

  \caption{Comparison of the regularized proposed scheme with V-BLAST modifications of Zero-Forcing and Babai approximation (for $N_{t}=4$ transmit antennas and $N_{r}=4$ single-antenna receivers with the rate $R=2$ bits per channel use per user.).}
  \label{fig:reduction-new}
\end{figure}

Figure 4 compares the performances of the fixed-rate and the
variable-rate transmission using lattice-basis reduction for $N_{t}=2$ transmit antennas and $N_{r}=2$ users. In both
cases, the sum-rate is 8 bits per channel use (in the case of fixed individual rates, a 16QAM constellation is assigned to each user). We see that by
eliminating the equal-rate constraint, we can considerably improve
the performance (especially, for high rates). In fact, the diversity
gains for the equal-rate and the unequal-rate methods are,
respectively, $ N_{r} $ and $ N_{t} N_{r}$.

\begin{figure}
  \centering
  \includegraphics[scale=.6,clip]{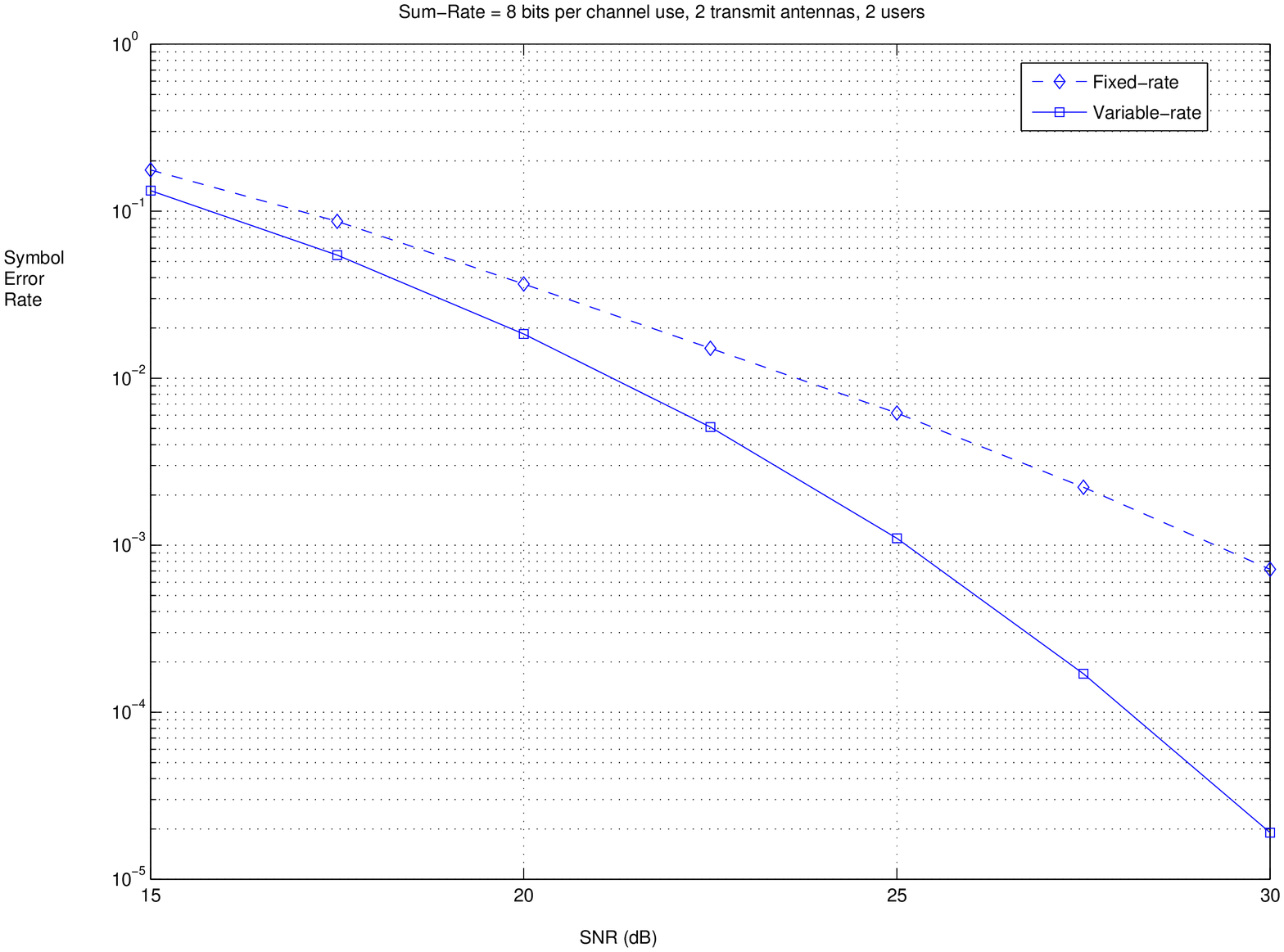}

  \caption{Performance comparison between the fixed-rate and the variable-rate transmission for $N_{t}=2$ transmit antennas and $N_{r}=2$ single-antenna receivers with sum-rate 8 bits per channel use.}
  \label{fig:reduction2}
\end{figure}

\section{Conclusion}

A simple scheme for communications in MIMO broadcast channels is
introduced which is based on the lattice reduction technique and
improves the performance of the channel inversion method. Lattice
basis reduction helps us to reduce the average transmitted energy by
modifying the region which includes the constellation points.
Simulation results show that the performance of the proposed scheme
is very close to the performance of the perturbation method. Also,
it is shown that by using lattice-basis reduction, we achieve the
maximum precoding diversity with a polynomial-time precoding
complexity.

\section*{Appendix A}

In this Appendix, we compute the second moment of a parallelotope
whose centroid is the origin and its edges are equal to
the basis vectors of the lattice.

Assume that $ \cal A $ is an $M$-dimensional  parallelotope
and $ X $ is its second moment. The second
moment of $ \frac{1}{2}\cal A $ is $ \left( \frac{1}{2}\right)
^{M+2}X $. The parallelotope $ \cal A $ can be considered as the
union of $ 2^{M} $ smaller parallelotopes which are constructed by
$ \pm\frac{1}{2} \mathbf{b}_{1}, \pm\frac{1}{2}
\mathbf{b}_{2},....,\pm \frac{1}{2}\mathbf{b}_{M} $, where $
\mathbf{b}_{i} $, $1 \leq i \leq M$, is a basis vector. These
parallelotopes are translated versions of $ \frac{1}{2}\cal A $
with the translation vectors $T_{i}=\pm\frac{1}{2} \mathbf{b}_{1}
\pm\frac{1}{2} \mathbf{b}_{2}\pm....\pm \frac{1}{2}\mathbf{b}_{M}
$, $1\leq i \leq 2^{M}$. The second moments of these
parallelotopes are equal to $ \left( \frac{1}{2}\right) ^{M+2}X +
\Vert T_{i} \Vert ^{2} \mbox{Vol}(\frac{1}{2}{\cal A})$, $1\leq i
\leq 2^{M}$. By the summation over all these second moments, we
can find the second moment of $ \cal A $.
\begin{equation}
%\begin{tabular}{c}
 X=\sum_{i=1}^{2^{M}}  \left[ \left( \frac{1}{2}\right) ^{M+2}X +
\Vert T_{i} \Vert ^{2}
.\mbox{Vol}(\frac{1}{2}{\cal A})\right]  \end{equation}

\begin{equation} = \left( \frac{1}{2}\right) ^{2}X+2^{M-2}(\Vert \mathbf{b}_{1}
\Vert^{2}+...+\Vert \mathbf{b}_{M}
\Vert^{2}).\mbox{Vol}(\frac{1}{2}{\cal A}) \end{equation}
\begin{equation} =  \frac{1}{4}X+ \frac{1}{4} (\Vert \mathbf{b}_{1}
\Vert^{2}+...+\Vert \mathbf{b}_{M} \Vert^{2}).\mbox{Vol}({\cal
A})\end{equation}
\begin{equation}\Longrightarrow X=\frac{1}{3}(\Vert \mathbf{b}_{1}
\Vert^{2}+...+\Vert \mathbf{b}_{M} \Vert^{2}).\mbox{Vol}({\cal
A}).
%\end{tabular}
\end{equation}

\section*{Appendix B}
\section*{Proof of Lemma 3}
Lemma 3 states that the probability that a lattice,  generated by
$ M $ independent $N $-dimensional complex Gaussian vectors, $N \geq M$, with a unit variance per each dimension,
has a nonzero point inside a sphere (centered at origin and with
the radius $ \varepsilon $) is bounded by $
\beta_{N,M}\varepsilon^{2N} $ for $N > M$, and $\beta_{N,M}\varepsilon^{2N} \max \left\lbrace (-\ln \varepsilon)^{N+1},1\right\rbrace $ for $ N=M\geq2$. We can assume that $ \varepsilon <1$ (for $ \varepsilon
\geq 1$, lemma 3 is trivial because the probability is bounded).

\subsection{Case 1: $M=1$}
When $M=1$, the lattice consists of the integer
multiples of the basis vector $\mathbf{v}$. If the norm of one of these vectors
is less than $\varepsilon$, then the norm of $\mathbf{v}$ is
less than $\varepsilon$. Consider the variance of the components of $\mathbf{v}$ as $ \varrho^{2}$. The vector $\mathbf{v}$ has an $N$-dimensional complex Gaussian distribution, $ f_{\mathbf{v}}(\mathbf{v})$. Therefore, the probability of this event is,
\begin{equation}\Pr\left\lbrace \Vert \mathbf{v} \Vert \leq \varepsilon\right\rbrace = \int_{\Vert \mathbf{v} \Vert \leq \varepsilon} f_{\mathbf{v}}(\mathbf{v}) \; d\mathbf{v} \leq \int_{\Vert \mathbf{v} \Vert \leq \varepsilon} \frac{1}{\pi^{N} \varrho^{2N}} \; d\mathbf{v} \leq \beta_{N,1}\frac{\varepsilon^{2N}}{\varrho^{2N}}. \end{equation}
When the variance of the components of $\mathbf{v}$ is equal to one, we have,
\begin{equation}\Pr\left\lbrace \Vert \mathbf{v} \Vert \leq \varepsilon\right\rbrace  \leq \beta_{N,1}\varepsilon^{2N}. \end{equation}

\subsection{Case 2: $N>M>1$}

Consider $ L_{(\mathbf{v}_{1},...,\mathbf{v}_{M})} $ as the lattice generated by $\mathbf{v}_{1} $,$\mathbf{v}_{2} $,...,$\mathbf{v}_{M}$. Each point of $ L_{(\mathbf{v}_{1},...,\mathbf{v}_{M})} $ can
be represented by $ \mathbf{v}_{(z_{1},...,z_{M})}=z_{1}\mathbf{v}_{1} + z_{2}\mathbf{v}_{2}+ ...+z_{M}\mathbf{v}_{M}$, where $ z_{1},...,z_{M}$ are complex integer numbers. The vectors $\mathbf{v}_{1} $,$\mathbf{v}_{2} $,...,$\mathbf{v}_{M}$ are independent and jointly Gaussian. Therefore, for every integer vector $ \mathbf{z}=(z_{1},...,z_{M})$, the entries of the vector $\mathbf{v}_{(z_{1},...,z_{M})}$ have complex Gaussian distributions with the variance 

\begin{equation}
\varrho_{\mathbf{z}}^{2}=\Vert \mathbf{z} \Vert^{2} \varrho^{2}= \left(\vert z_{1}\vert^{2}+...+\vert z_{M}\vert^{2} \right) \varrho^{2} .
\end{equation}
Therefore, according to the lemma for $M=1$, 
\begin{equation}\Pr\left\lbrace \Vert \mathbf{v}_{(z_{1},...,z_{M})} \Vert \leq \varepsilon\right\rbrace  \leq \beta_{N,1}\frac{\varepsilon^{2N}}{\left(\vert z_{1}\vert^{2}+...+\vert z_{M}\vert^{2} \right)^{N} }. \end{equation}
Now, by using the union bound,

\begin{equation}\Pr\left\lbrace d_{\mathbf{H}} \leq \varepsilon\right\rbrace \leq \sum_{\mathbf{z}\neq0}\Pr\left\lbrace \Vert \mathbf{v}_{(z_{1},...,z_{M})} \Vert \leq \varepsilon\right\rbrace   \end{equation}

\begin{equation}\leq  \sum_{\mathbf{z}\neq0} \beta_{N,1}\frac{\varepsilon^{2N}}{\left(\vert z_{1}\vert^{2}+...+\vert z_{M}\vert^{2} \right)^{N} }
\end{equation}
$$=  \beta_{N,1} \left( \sum_{1\leq \Vert\mathbf{z}\Vert <2}\frac{\varepsilon^{2N}}{\Vert \mathbf{z}\Vert^{2N} } +\sum_{2\leq \Vert\mathbf{z}\Vert <3} \frac{\varepsilon^{2N}}{\Vert \mathbf{z}\Vert^{2N} } \right.
$$

\begin{equation}\label{eq:92} \left. +  \sum_{3\leq \Vert\mathbf{z}\Vert <4} \frac{\varepsilon^{2N}}{\Vert \mathbf{z}\Vert^{2N} } + ...\right) .
\end{equation}

The $M$-dimensional complex integer points $\mathbf{z}=(z_{1},...,z_{M}) $, such that $ k\leq \Vert\mathbf{z}\Vert <k+1$, can be considered as the centers of disjoint unit-volume cubes. All these cubes are inside the region between the $2M$-dimensional spheres, with radii $k-1$ and $k+2$. Therefore, the number of $M$-dimensional complex integer points $\mathbf{z}=(z_{1},...,z_{M}) $, such that $ k\leq \Vert\mathbf{z}\Vert <k+1$, can be bounded by the volume of the region between these two $2M$-dimensional spheres. Thus, this number is bounded by $ c_{1} k^{2M-1}$ for some constant\footnote{Throughout this
proof, $c_{1},c_{2},... $ are some constant numbers.} $ c_{1} $. Therefore,
\begin{equation} \label{eq:95} \sum_{k\leq \Vert\mathbf{z}\Vert <k+1} \frac{\varepsilon^{2N}}{\Vert\mathbf{z}\Vert^{2N} } \leq c_{1} k^{2M-1} \frac{\varepsilon^{2N}}{k^{2N} }
\end{equation}
$$ {\rm (\ref{eq:92}), (\ref{eq:95})} \Longrightarrow \Pr\left\lbrace d_{\mathbf{H}} \leq \varepsilon\right\rbrace \leq c_{1} \beta_{N,1}\varepsilon^{2N} + 2^{2M-1}c_{1} \beta_{N,1}\frac{\varepsilon^{2N}}{2^{2N} }+ $$

\begin{equation} + 3^{2M-1}c_{1} \beta_{N,1}\frac{\varepsilon^{2N}}{3^{2N} } + ...
\end{equation}

\begin{equation} \leq c_{1}\beta_{N,1}\varepsilon^{2N} \sum_{k=1}^{\infty} \frac{1}{k^{2N-2M+1}}.
\end{equation}

According to the assumption of this case, $N>M$; hence, $2N-2M+1 \geq 2 $. Therefore, the above summation is convergent:

\begin{equation}\Pr\left\lbrace \Vert \mathbf{v} \Vert \leq \varepsilon\right\rbrace \leq \beta_{N,M}\varepsilon^{2N} .
\end{equation}

\subsection{Case 3: $N=M>1$}

Each point of $ L_{(\mathbf{v}_{1},...,\mathbf{v}_{N})} $ can
be represented by $ z\mathbf{v}_{N}-\mathbf{v} $, where $\mathbf{v} $ belongs to the lattice $ L_{(\mathbf{v}_{1},...,\mathbf{v}_{N-1})} $ and $z$ is a complex integer. 
Consider $\mathcal{S}_{\mathbf{v}} $ as the sphere with radius $ \varepsilon $ and centered at ${\mathbf{v}}$. Now, $ z\mathbf{v}_{N}-\mathbf{v} $ belongs to $\mathcal{S}_{\mathbf{0}} $ iff the $ z\mathbf{v}_{N} $ belongs to $\mathcal{S}_{\mathbf{v}} $. 
Also, the sphere $\mathcal{S}_{\mathbf{v}} $ includes a
point $ z\mathbf{v}_{N}$ iff $ \mathcal{S}_{\mathbf{v},z} $
includes $ \mathbf{v} $ , where $ \mathcal{S}_{\mathbf{v},z}=\frac{ \mathcal{S}_{\mathbf{v}}}{z} $ is the
sphere centered at $ \mathbf{v}/z $ with radius
$\dfrac{\varepsilon}{\vert z \vert} $ (see figure 4). Therefore, the probability
that a lattice point exists in $ \mathcal{S}_{\mathbf{v}} $ is equal to the
probability that $ \mathbf{v}_{N} $ is in at least one of the
spheres $ \left\lbrace \mathcal{S}_{\mathbf{v},z} \right\rbrace $, $z\neq 0$.
 
 \begin{figure}
  \centering
  \includegraphics[scale=.5,clip]{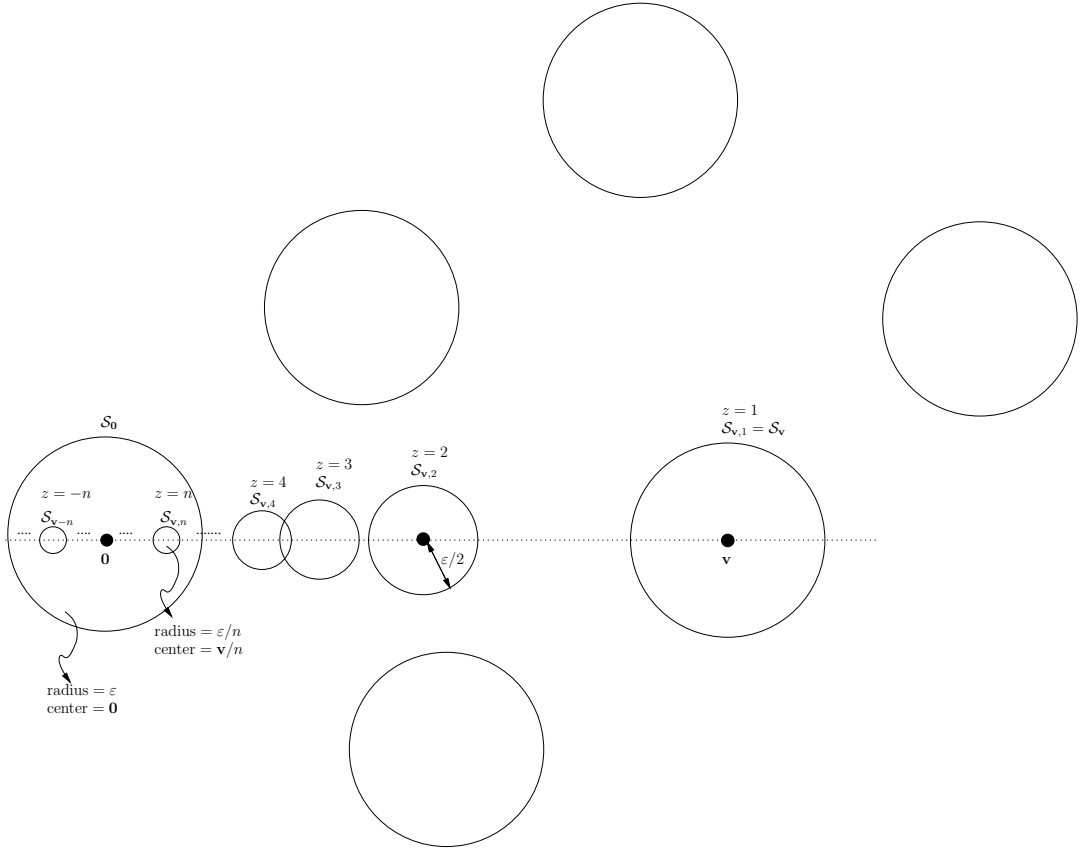}

  \caption{The family of spheres $ {\mathcal{S}_{\mathbf{v},z}}$}
  \label{fig:proof}
\end{figure}
 
 If we consider $d_{\mathbf{H}}$ as the minimum distance of $ L_{(\mathbf{v}_{1},...,\mathbf{v}_{N})} $ and $R$ as an arbitrary number greater than 1:

\begin{equation} \label{eq:100}    
\Pr\left\lbrace  d_{\mathbf{H}}  \leq \varepsilon \right\rbrace = \Pr \left\lbrace\left(  L_{(\mathbf{v}_{1},...,\mathbf{v}_{N})}-0\right) \cap \mathcal{S}_{\mathbf{0}} \neq \varnothing \right\rbrace = \Pr \left\lbrace \mathbf{v}_{N} \in \bigcup_{\mathbf{v}}\bigcup_{z\neq0} \mathcal{S}_{\mathbf{v},z} \right\rbrace \end{equation}

\begin{equation} \label{eq:101} 
\leq \Pr \left\lbrace \mathbf{v}_{N} \in \bigcup_{\Vert \frac{\mathbf{v}}{z}
 \Vert \leq R}  \mathcal{S}_{\mathbf{v},z} \right\rbrace + \Pr \left\lbrace \mathbf{v}_{N} \in \bigcup_{\Vert \frac{\mathbf{v}}{z} \Vert>R } \mathcal{S}_{\mathbf{v},z} \right\rbrace
\end{equation}
In the second term of (\ref{eq:101}), all the spheres have centers with norms greater than $ R$ and radii less than 1  (because $ \vert z \vert \geq 1$). Therefore,
\begin{equation} \label{eq:102} 
 \bigcup_{\Vert \frac{\mathbf{v}}{z} \Vert>R } \mathcal{S}_{\mathbf{v},z} \subset \left\lbrace \mathbf{x} \left\vert  \right. \Vert \mathbf{x} \Vert > R-1 \right\rbrace
\end{equation}

\begin{equation}
{\rm (\ref{eq:102})} \Longrightarrow {\rm (\ref{eq:101})}
\leq \Pr \left\lbrace \mathbf{v}_{N} \in \bigcup_{\Vert \frac{\mathbf{v}}{z}
 \Vert \leq R}  \mathcal{S}_{\mathbf{v},z} \right\rbrace + \Pr \left\lbrace \Vert \mathbf{v}_{N} \Vert > R-1 \right\rbrace
\end{equation}

\begin{equation} \label{eq:104}
\leq \Pr \left\lbrace \mathbf{v}_{N} \in \bigcup_{\Vert \frac{\mathbf{v}}{z}
 \Vert \leq R, \vert z \vert \leq \varepsilon^{1-N}}  \mathcal{S}_{\mathbf{v},z} \right\rbrace +  \Pr \left\lbrace \mathbf{v}_{N} \in \bigcup_{\Vert \frac{\mathbf{v}}{z}
 \Vert \leq R, \vert z
 \vert > \varepsilon^{1-N}}  \mathcal{S}_{\mathbf{v},z} \right\rbrace + \Pr \left\lbrace \Vert \mathbf{v}_{N} \Vert > R-1 \right\rbrace.
\end{equation}
We bound the first term of (\ref{eq:104}) as the following:
\begin{equation} \label{eq:105}
 \Pr\lbrace \mathbf{v}_{N} \in \bigcup_{\Vert \frac{\mathbf{v}}{z}
 \Vert \leq R, \vert z \vert \leq \varepsilon^{1-N}}  \mathcal{S}_{\mathbf{v},z} \rbrace 
\end{equation}

$$
\leq \left(  \sum_{\Vert \mathbf{v} \Vert \leq 2R} \sum_{\vert z \vert \geq 1} \Pr\lbrace \mathbf{v}_{N} \in \mathcal{S}_{\mathbf{v},z} \rbrace +  \sum_{2R<\Vert \mathbf{v} \Vert \leq 3R} \sum_{\vert z \vert \geq 2}\Pr\lbrace \mathbf{v}_{N} \in \mathcal{S}_{\mathbf{v},z} \rbrace + \right.
$$

\begin{equation}
\left.  ... + \sum_{\lfloor \varepsilon^{1-N} \rfloor R <\Vert \mathbf{v} \Vert \leq (\lfloor \varepsilon^{1-N} \rfloor+1)R} \;\;\; \sum_{\vert z \vert \geq \lfloor \varepsilon^{1-N} \rfloor} \Pr\lbrace \mathbf{v}_{N} \in \mathcal{S}_{\mathbf{v},z} \rbrace \right).
\end{equation}
Noting that the pdf of $ \mathbf{v}_{N}$ is less than or equal to $ \frac{1}{\pi^{N}}$,
$$
{\rm (\ref{eq:105})} 
\leq  \frac{1}{\pi^{N}}\left(  \sum_{\Vert \mathbf{v} \Vert \leq 2R} \sum_{\vert z \vert \geq 1} {\rm Vol} (\mathcal{S}_{\mathbf{v},z} ) +  \sum_{2R<\Vert \mathbf{v} \Vert \leq 3R} \sum_{\vert z \vert \geq 2}{\rm Vol} (\mathcal{S}_{\mathbf{v},z} ) + \right.
$$

\begin{equation}
\left.  ... + \sum_{\lfloor \varepsilon^{1-N} \rfloor R <\Vert \mathbf{v} \Vert \leq (\lfloor \varepsilon^{1-N} \rfloor+1)R} \;\;\; \sum_{\vert z \vert \geq \lfloor \varepsilon^{1-N} \rfloor}  {\rm Vol} (\mathcal{S}_{\mathbf{v},z} ) \right)
\end{equation}

$$
\leq  \frac{1}{\pi^{N}}\left(  \sum_{\Vert \mathbf{v} \Vert \leq 2R} \sum_{\vert z \vert \geq 1} \frac{c_{2}\varepsilon^{2N}}{\vert z \vert ^{2N}} +  \sum_{2R<\Vert \mathbf{v} \Vert \leq 3R} \sum_{\vert z \vert \geq 2}\frac{c_{2}\varepsilon^{2N}}{\vert z \vert ^{2N}} + \right.
$$

\begin{equation}
\left. ... + \sum_{\lfloor \varepsilon^{1-N} \rfloor R <\Vert \mathbf{v} \Vert \leq (\lfloor \varepsilon^{1-N} \rfloor+1)R} \;\;\; \sum_{\vert z \vert \geq \lfloor \varepsilon^{1-N} \rfloor} \frac{c_{2}\varepsilon^{2N}}{\vert z \vert ^{2N}}  \right).
\end{equation}
By using (\ref{eq:95}), for one-dimensional complex vector $\mathbf{z}=z $,
\begin{equation} \label{eq:110}
\sum_{\vert z \vert \geq i} \frac{c_{2}\varepsilon^{2N}}{\vert z \vert ^{2N}} = \sum_{k=i}^{\infty} c_{2} .\sum_{k \leq \vert z \vert \leq k+1}\frac{\varepsilon^{2N}}{\vert z \vert ^{2N}} \leq \sum_{k=i}^{\infty} \frac{c_{1}c_{2}\varepsilon^{2N}}{k ^{2N-1}} \leq  \frac{c_{3}\varepsilon^{2N}}{i ^{2N-2}} 
\end{equation}
Now,
$$
{\rm (\ref{eq:110})} \Longrightarrow {\rm (\ref{eq:105})}
\leq  \frac{1}{\pi^{N}} \left(  \sum_{\Vert \mathbf{v} \Vert \leq 2R}  c_{3}\varepsilon^{2N}+  \sum_{2R<\Vert \mathbf{v} \Vert \leq 3R} \frac{c_{3}\varepsilon^{2N}}{2^{2N-2}} + \right.
$$

\begin{equation} \label{eq:153}
\left. ... + \sum_{\lfloor \varepsilon^{1-N} \rfloor R <\Vert \mathbf{v} \Vert \leq (\lfloor \varepsilon^{1-N} \rfloor+1)R}  \frac{c_{3}\varepsilon^{2N}}{\lfloor \varepsilon^{1-N} \rfloor^{2N-2}}  \right).
\end{equation}

Assume that the minimum distance of  $ L_{(\mathbf{v}_{1},...,\mathbf{v}_{N-1})} $ is $d_{N-1}$. The spheres with the radius $d_{N-1}/2$ and centered by the points of $ L_{(\mathbf{v}_{1},...,\mathbf{v}_{N-1})} $ are disjoint. Therefore, the number of points from the ($N-1$)-dimensional complex lattice $ L_{(\mathbf{v}_{1},...,\mathbf{v}_{N-1})} $, such that $\Vert \mathbf{v} \Vert \leq 2R$, is bounded by  $\frac{c_{4}(2R+d_{N-1}/2)^{2N-2}}{d_{N-1}^{2N-2}} $ (it is bounded by the ratio between the volumes of ($2N-2$)-dimensional spheres with radii $2R+d_{N-1}/2$ and $d_{N-1}/2 $). Also, the number of points from $ L_{(\mathbf{v}_{1},...,\mathbf{v}_{N-1})} $, such that $(k-1)R<\Vert \mathbf{v} \Vert \leq kR$, is bounded by $\frac{c_{4}(kR)^{2N-3}(R+d_{N-1})}{d_{N-1}^{2N-2}} $ (it is bounded by the ratio between the volumes of the region defined by $ (k-1)R-d_{N-1}/2<\Vert \mathbf{x} \Vert \leq kR+d_{N-1}/2$ and the sphere with radius $d_{N-1}/2 $):

\begin{equation} (\ref{eq:153})
\leq   \frac{c_{5}(2R+d_{N-1}/2)^{2N-2}}{d_{N-1}^{2N-2}}.\varepsilon^{2N} + \frac{c_{5}R^{2N-3}(R+d_{N-1})}{d_{N-1}^{2N-2}}.\varepsilon^{2N} \sum_{k=2}^{\lfloor \varepsilon^{1-N} \rfloor} \frac{1}{k} 
\end{equation}

\begin{equation} (\ref{eq:153})
\leq   \frac{c_{5}(2R+d_{N-1}/2)^{2N-2}}{d_{N-1}^{2N-2}}.\varepsilon^{2N} + \frac{c_{5}R^{2N-3}(R+d_{N-1})}{d_{N-1}^{2N-2}}.\varepsilon^{2N} . \ln (\varepsilon^{1-N})  
\end{equation}

\begin{equation} \label{eq:156}
\leq  c_{6}\varepsilon^{2N}. \max \left(\frac{R^{2N-2}}{d_{N-1}^{2N-2}} ,1 \right). \max \left\lbrace - \ln \varepsilon,1 \right\rbrace .
\end{equation}

According to the proof of the case 2, we have $ \Pr\left\lbrace d_{N-1} \leq \eta \right\rbrace \leq \beta_{N,N-1} \eta^{2N} $. Therefore, 

\begin{equation}
{\rm E}_{d_{N-1}} \left\lbrace \max \left(\frac{R^{2N-2}}{d_{N-1}^{2N-2}} ,1 \right) \right\rbrace 
\end{equation}

\begin{equation}
\leq 1 . \Pr \left\lbrace d_{N-1} >R \right\rbrace+ 2^{2N-2}. \Pr \left\lbrace \frac{1}{2}R< d_{N-1} \leq R  \right\rbrace
+3^{2N-2} . \Pr \left\lbrace \frac{1}{3}R< d_{N-1} \leq \frac{1}{2}R  \right\rbrace + ...
\end{equation}

\begin{equation}
\leq 1 + 2^{2N-2}. \Pr \left\lbrace d_{N-1} \leq R  \right\rbrace
+3^{2N-2} . \Pr \left\lbrace  d_{N-1} \leq \frac{1}{2}R  \right\rbrace + ...
\end{equation}

\begin{equation}
\leq 1 + \sum_{k=1}^{\infty} \frac{(k+1)^{2N-2}}{k^{2N}}. R^{2N}\beta_{N,N-1}
\leq c_{7}R^{2N}
\end{equation}
\begin{equation} \label{eq:123}
\Longrightarrow {\rm E}_{d_{N-1}}\left\lbrace c_{6}\varepsilon^{2N}. \max \left(\frac{R^{2N-2}}{d_{N-1}^{2N-2}} ,1 \right). \max \left\lbrace - \ln \varepsilon,1 \right\rbrace \right\rbrace \leq c_{8} \varepsilon^{2N}. R^{2N}. \max \left\lbrace - \ln \varepsilon,1 \right\rbrace.
\end{equation}

To bound the second term of (\ref{eq:104}), we note that for $ \vert z \vert \geq \varepsilon^{1-N} $, the radii of the spheres $\mathcal{S}_{\mathbf{v},z}$ are less or equal to $ \varepsilon^{N}$, and the centers of these spheres lie on the $(N-1)$-dimensional complex subspace containing $L_{\mathbf{v}_{1},...,\mathbf{v}_{N-1}}$. Also, the norm of these centers are less than $ R$. Therefore, all of these spheres are inside the region $ \mathcal{A} $ which is an orthotope centered at the origin, with $2N$ real dimensions (see figure 5): 

\begin{equation}
 \bigcup_{\Vert \frac{\mathbf{v}}{z}
 \Vert \leq R, \vert z
 \vert > \varepsilon^{1-N}}  \mathcal{S}_{\mathbf{v},z} \subset \mathcal{A}
\end{equation}

\begin{equation}
\Longrightarrow \Pr\lbrace \mathbf{v}_{N} \in \bigcup_{\Vert \frac{\mathbf{v}}{z}
 \Vert \leq R, \vert z
 \vert > \varepsilon^{1-N}}  \mathcal{S}_{\mathbf{v},z} \rbrace \leq \frac{1}{\pi^{N}} {\rm Vol}\left( \bigcup_{\Vert \frac{\mathbf{v}}{z}
 \Vert \leq R, \vert z
 \vert > \varepsilon^{1-N}}  \mathcal{S}_{\mathbf{v},z}\right) \leq \frac{1}{\pi^{N}}{\rm Vol}(\mathcal{A})
\end{equation}

\begin{equation} \label{eq:125}
\leq \frac{1}{\pi^{N}} (2 \varepsilon^{N})^{2}(2R+\varepsilon^{N})^{2N-2} 
\end{equation}

\begin{figure}
  \centering
  \includegraphics[scale=.5,clip]{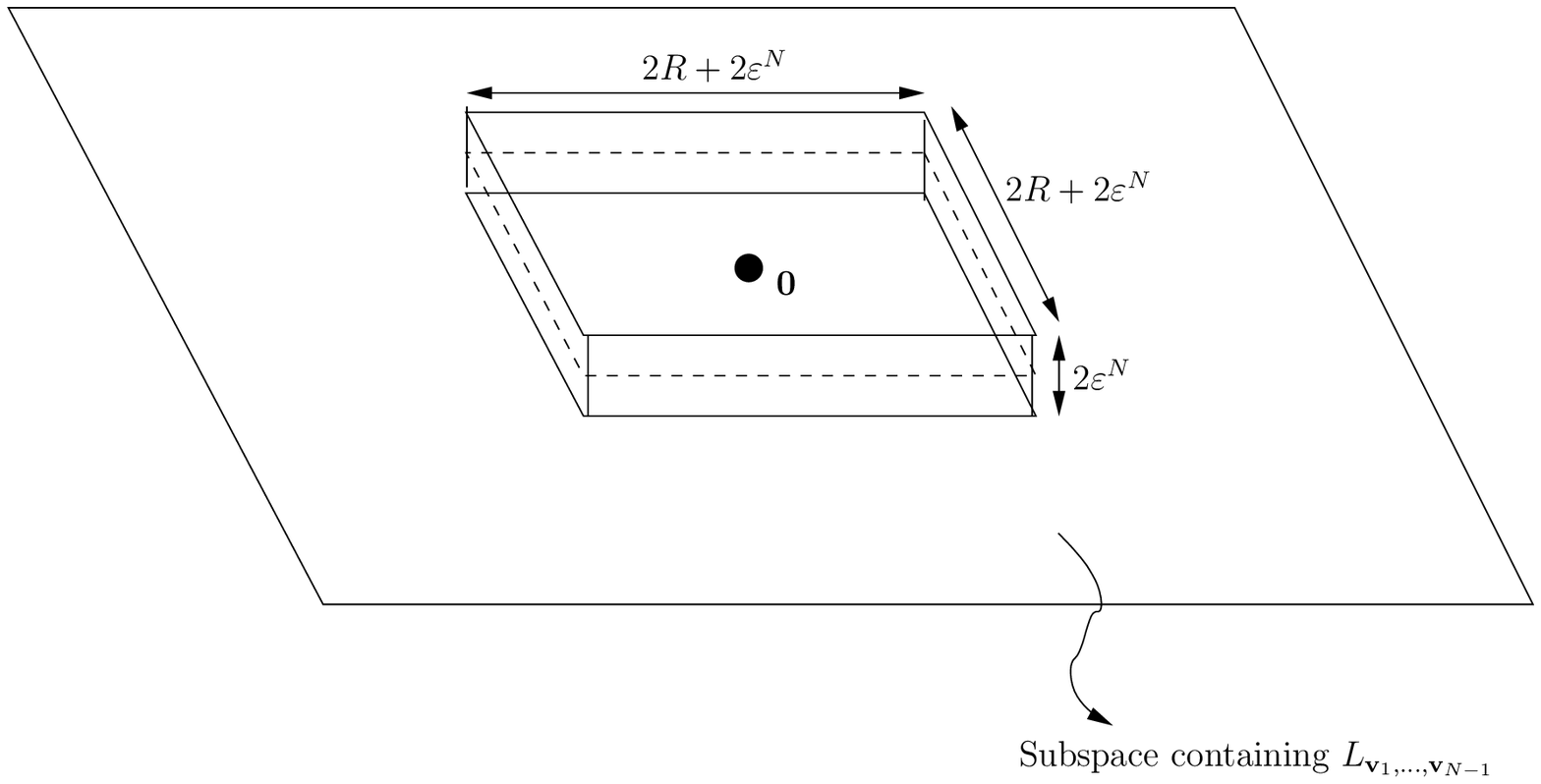}

  \caption{The orthotope $\mathcal{A}$}
  \label{fig:proof2}
\end{figure}

Also, according to the Gaussian distribution of the entries of $\mathbf{v}_{N}$ (which have Variance $ \frac{1}{2}$ on each real dimension), we can bound the third term of (\ref{eq:104}) as,
\begin{equation} \label{eq:126}
\Pr\lbrace \Vert \mathbf{v}_{N} \Vert > R-1 \rbrace \leq  2N Q \left( \sqrt{\frac{R-1}{N}} \right)   \leq c_{9}e^{-\left(\frac{R-1}{\sqrt{2}N}\right)^{2}}.
\end{equation}
By using (\ref{eq:123}), (\ref{eq:125}), and (\ref{eq:126}),

\begin{equation}
 \Pr\left\lbrace   d_{\mathbf{H}} \leq \varepsilon \right\rbrace  \leq c_{8} \varepsilon^{2N}. R^{2N}. \max \left\lbrace - \ln \varepsilon,1 \right\rbrace + \frac{1}{\pi^{N}} (2\varepsilon^{N})^{2}(2R+\varepsilon^{N})^{2N-2}+c_{9}e^{-\left(\frac{R-1}{\sqrt{2}N}\right)^{2}}.\end{equation}
The above equation is true for every $R>1$. Therefore, using $R=\sqrt{2}N\sqrt{-\ln(\varepsilon^{2N})}+1$,
\begin{equation}
\Pr\left\lbrace   d_{\mathbf{H}} \leq \varepsilon \right\rbrace \leq
\beta_{N,N}\varepsilon^{2N}.\max \left\lbrace (-\ln \varepsilon)^{N+1},1 \right\rbrace.
\end{equation}

\end{document}